\documentclass[]{aa}  
\usepackage{graphicx}
\usepackage{txfonts}
\usepackage{xcolor}
\usepackage{amsmath}
\usepackage{multirow}
\usepackage{lipsum}

\usepackage[colorlinks=true,
            linkcolor=blue,
            citecolor=blue,
            filecolor=black,
            urlcolor=blue]{hyperref}

\begin{document}

   \title{Radio spectral properties and aging of two tailed radio galaxies in a galaxy group at z=0.35}
   
   \author{P. Vuli\'c
          \inst{1}\fnmsep\thanks{Email: pvulic@phy.hr}
          \and
          V. Smol\v{c}i\'c\inst{1}
          \and
          A. Finoguenov\inst{2}
          \and
          G. Gozaliasl\inst{2,3}
          \and
          H.S.B. Algera\inst{4}
          \and
          I. Delvecchio\inst{5}
          }

   \institute{Department of Physics, University of Zagreb, Bijeni\v{c}ka cesta 32, Zagreb, Croatia
        \and
            Department of Physics, University of Helsinki, P.O. Box 64, FI-00014 Helsinki, Finland
        \and
            Department of Computer Science, Aalto University, PO Box 15400, 00076, Espoo, Finland
        \and
            Institute of Astronomy and Astrophysics, Academia Sinica, 11F of Astronomy-Mathematics Building, No.1, Sec. 4, Roosevelt Rd, Taipei 106319, Taiwan, R.O.C.
        \and
            INAF – Osservatorio di Astrofisica e Scienza dello Spazio di Bologna, Via Gobetti 93/3, I-40129 Bologna, Italy
            }
   \date{Received May 28 2026; accepted ...}
 
\abstract
{
We present a study of two tailed radio galaxies located in the core of a massive, dynamically young galaxy group, previously identified as an early group-group merger. Using broad-band VLA 3 GHz and 1.4 GHz, MeerKAT 1.35 GHz, and GMRT 610 MHz and 325 MHz observations, we investigate the radio spectral properties as well as the spectral and dynamical ages of both sources.
The radio morphologies show clear evidence of interaction with the intragroup medium (IGM). One galaxy is a wide-angle tail (WAT) source with well-collimated, Mpc-scale bent jets terminating in bright lobes, while the other is most likely a head–tail (HT) galaxy characterized by a compact core and a mostly straight tail that bends slightly near its end. Both galaxies exhibit high radio luminosities across all five frequencies, and we find spectral indices of $\alpha=0.8\pm 0.1$ for the WAT and $\alpha=0.6\pm 0.2$ for the HT galaxy. Spatially resolved spectral index maps and spectral analysis of individual galaxy components reveal steepening of the radio spectrum with increasing distance from the core in both galaxies, with localized spectral flattening in the WAT lobes and hotspots along the northern jet, and indications of such flattening in the middle of the HT tail.
Spectral ages derived using the Jaffe-Perola model are $33.80\substack{+7.63 \\ -7.23}$ Myr for the WAT and $20.86\substack{+10.07 \\ -17.17}$ Myr for the HT, significantly lower than the corresponding dynamical age estimates of $420\pm60$ to $700\pm100$ Myr (for WAT) and $140\pm20$ Myr (and possibly up to $280\pm40$ Myr, for HT), yielding dynamical-to-spectral age ratios of $\sim12-20$ and $\sim7$ (and possibly up to $\sim14$), respectively. The discrepancy may be reduced in future studies through the use of more complex dynamical aging models that incorporate interactions between the radio lobes and the IGM, which require deeper X-ray observations of the group. Spectral age estimates may also be affected by mixing of electron populations, and could be better constrained with future deep, high-resolution broad-band radio observations of this system at both MHz and frequencies above 3 GHz.
The combination of extended radio structures, spectral signatures of radiative aging with localized re-acceleration, and activity timescales of tens to hundreds of Myr indicates that both galaxies are actively interacting with, and likely mechanically depositing energy into their environment.
}

   \maketitle

\section{Introduction}
\label{sec:intro}
Radio galaxies are often found in dense environments such as galaxy clusters and groups \citep{best2005}. These sources typically eject powerful jets of plasma into their environment, inflating radio lobes, and as widely proposed, they can affect their host galaxies through AGN-feedback (\citealt{croton2006}, \citealt{bower2006}, also see \citealt{fabian2012} for a review of observational evidence). There is evidence for powerful jets significantly heating the interstellar medium (ISM) of the host galaxy \citep{kraft2003,kraft2007,croston_kraft_hardcastle2007,croston2009}. This could in principle affect (suppress) the galaxy's star formation, acting as a self-regulating mechanism in galaxy mass growth, and therefore explain the observed decrease at the high-mass end of the galaxy mass function \citep{croton2006,bower2006}. For the most powerful radio galaxies, the jets may traverse the host galaxy and terminate in the intracluster or the intragroup medium (ICM or IGM), inducing shocks, inflating cavities, and affecting its thermodynamical properties \citep{birzan2004,birzan2008,liu2019,liu2020}. These powerful radio sources are typically hosted by massive elliptical galaxies, corresponding to the brightest cluster or the brightest group galaxy (BCG or BGG, \citealt{best2007}, \citealt{croft2007}). In general, the heating and possible mixing of the material can affect the evolution of individual host galaxies, but also the evolution of the hosting galaxy clusters or groups.\par
Studying morphology and radio spectral properties of jets and lobes provides insights into the physics of jet launching, activity and evolution, and can also reflect the dynamical state of the local environment. Radio galaxies are commonly divided into FRI and FRII sources \citep{FR1974}. In the former, radio brightness decreases from the core towards the source's edge, making them edge-darkened, and spectral index increases, implying aging of the electron population. The latter, on the contrary, exhibit bright hotspots on their edges (in the lobes). Most powerful radio galaxies are typically FRII sources \citep{FR1974}. In these sources spectral index steepens with the distance from the core along the jets, however it flattens again in the radio lobes, probably due to particle re-acceleration in the hotspots. Bent radio morphologies are commonly observed in dynamically young galaxy clusters and groups \citep{smolcic2007,oklopcic2010,douglass2011,dasadia2016}, as a consequence of interaction of the radio galaxy structure, in particular the jets, with the surrounding medium \citep{begelman1979}. Bent radio galaxies are typically categorized based on the opening angle between their jets. When the jets exhibit sharp, collimated bending, the sources are classified as narrow-angle tail (NAT) galaxies \citep{miley1972,RO1976}, while those with broad jet angles are known as wide-angle tail (WAT) galaxies \citep{OR1976}. In cases where the individual jets cannot be clearly distinguished, these sources are often grouped more generally under the term head-tail (HT) galaxies. The latter can appear due to a number of reasons, i.e. due to a strong bending, limited resolution, projection effects, or a combination of these.\par
Understanding and quantifying the energetics of radio galaxies during their outburst phase, and estimating its duration are crucial to better constrain the physics of feedback (heating) mechanisms and their effect on cosmological evolution. Radio sources' age estimates are commonly used in quantifying the energetics, as the jet power can, in principle, be inferred by dividing the total energy output of the source by its age \citep{birzan2004,ineson2017}. However, values resulting from radiative aging analysis are regularly in disagreement with dynamical ages from different evolution models, i.e. spectral ages are reported to be multiple times smaller than dynamical ages, known as the age discrepancy problem or spectral age problem \citep{harwood2013, harwood_hardcastle_croston2015, mahatma2020}. The origin of age differences is still not well understood, however, mixing of electron populations is suspected to underestimate spectral ages \citep{harwood2016,turner2017}, while some of the dynamical age calculation methods in the literature \citep{birzan2004, ineson2017}, assuming constant lobe advance speed, could be overestimating the true age. Furthermore, multiple approaches have been proposed to estimate the total energy content of radio sources \citep[e.g.][]{birzan2004,rafferty2006,ineson2017}, and it is important to note that alternative jet power estimators exist and are sometimes adopted in the literature \citep{lobanov1998,GS2013}.\par

In this paper, we analyze radio morphology, spectral properties, and aging of two tailed radio galaxies found in a massive, dynamically young, galaxy group in the \textit{Cosmic Evolution Survey} (COSMOS, \citealt{cosmos})\footnote{\url{https://cosmos.astro.caltech.edu/page/astronomers}} field at $z=0.349$.\par
The group was previously detected in X-ray as one of the 247 galaxy groups identified in the COSMOS field by \citet{Gozaliasl}, studied among 10 other groups in COSMOS by \citet{Kettula}, and a detailed study of its structure and dynamics was performed using a multiwavelength dataset in \citet{Vulic2025}.
The results of \citet{Vulic2025} suggest this group is an unrelaxed system, likely an early stage merger of smaller structures (subgroups) of galaxies. An asymmetric and irregular distribution of both the member galaxies and the IGM was found, the latter exhibiting a high temperature (derived from X-ray spectra) $T_X=2.4\pm0.6$~keV and an electron density of $(8.2\pm0.3) \times 10^{-4}\hspace{0.1cm}\mathrm{cm^{-3}}$. High velocity bulk motion of the IGM was detected in the group core, with $v_{IGM}\gtrsim380\hspace{0.1cm}\mathrm{km/s}$ \citep{Vulic2025}. The group is very luminous, massive, and hot, with $L_X$ $\approx 750\%$, $M_{200}$ $\approx 320\%$, and $T_X$ $\approx 130\%$ above the corresponding medians for X-ray detected galaxy groups in the COSMOS field in the redshift bin [0.1,0.5] (\citealt{Vulic2025}, based on the data of \citealt{Gozaliasl}). The BGG and the second brightest group galaxy host the two tailed radio galaxies we study in this work. For more details we refer the reader to \citet{Vulic2025}.\par
Here, in Section~\ref{sec:data} we introduce the radio data we use. In Section~\ref{sec:tailed_radio_morph} we analyze the spectral evolution of radio galaxies' morphology, in Section~\ref{sec:integrated_flux_density_spectra} we find integrated flux densities, luminosites, and spectral indices, and in Section~\ref{sec:specindmaps} we generate spectral index maps. In Section~\ref{sec:age_analysis} we estimate spectral and dynamical ages. Discussion is given in Section~\ref{sec:discussion} and a short summary in Section~\ref{sec:summary}. Throughout the paper, we refer to the two tailed radio galaxies as 10913 (WAT) and 44 (HT), according to their 3 GHz catalog IDs (see Section ~\ref{sec:data} for references), and following the nomenclature of \citet{Vulic2025}.\par
Throughout the paper, we use $H_0=69.32\hspace{0.15cm}\mathrm{km\hspace{0.1cm}Mpc^{-1}\hspace{0.1cm}s^{-1}}$, $\Omega_M=0.2865$, and $\Omega_{\lambda}=0.7135$, and we define the radio spectral index to be positive, namely $F_\nu\propto \nu^{-\alpha}$.
\vspace{-0.2cm}
\section{Data}

\label{sec:data}
To conduct the analysis we use flux density maps (or mosaics), the corresponding rms noise maps, and (if available) radio source catalogs, in the COSMOS field, at 5 different frequencies (7 different data sets). These were generated from Very Large Array (VLA): 3 GHz, 3 GHz XS (Deep), 1.4 GHz (Large and Deep);
MeerKAT telescope: 1.282 GHz (1.35 GHz effectively); and Giant Meter Radio Telescope (GMRT): 610 MHz and 325 MHz observations. See below for references.

We use data from the VLA-COSMOS 3 GHz Large Project \citep{3ghz_smol_a}. This includes the 3 GHz continuum mosaic (hereafter 3 GHz map), generated from 384 h of VLA observations toward the $2\hspace{0.1cm}\mathrm{deg}^2$ COSMOS field, and characterized by a $0.75^{\prime\prime}$ resolution, and an average $2.3\hspace{0.1cm}\mathrm{\mu Jy/beam}$ rms noise. We also use the corresponding rms noise map and radio source catalog \citep{3ghz_smol_a}, along with its extended version, COSMOS VLA 3 GHz Multiwavelength Counterpart catalog (\citealt{3ghz_smol_b}, hereafter VLA 3 GHz catalog).

We also use data from the Ultradeep Multiband VLA Survey of the Faint Radio Sky \citep[hereafter COSMOS-XS,][]{vanderVlugt2021}. They observed $\sim 180\hspace{0.1cm}\mathrm{arcmin^2}$ of the COSMOS field at 3 GHz (S band) and $\sim 16\hspace{0.1cm}\mathrm{arcmin^2}$ at 10 GHz (X band). Here, we only use the 3 GHz radio map and the corresponding radio source catalog \citep[hereafter 3 GHz Deep map and catalog,][]{vanderVlugt2021}, as our targets lie outside the 10 GHz coverage area. The 3 GHz Deep map was produced from a single deep pointing with a total integration time of 100 h, and reaches a resolution of $\sim 2^{\prime\prime}$, and median rms 0.53 $\mathrm{\mu Jy\hspace{0.1cm}beam^{-1}}$. At the location of the radio galaxies, the primary beam sensitivity corresponds to $\sim40\%$ of its peak value, resulting in a $\sim2.5\times$ higher rms than at the pointing center. At 3 GHz, this survey is $\sim 5 \times$ deeper than the survey of \citet{3ghz_smol_a}. For this paper, we construct the corresponding rms noise map from the primary beam corrected 3 GHz Deep map using AIPS RMSD 100 pixels circular mesh.\par

We use 1.4 GHz data from the Large and Deep projects of the VLA-COSMOS Survey \citep{Schinnerer_L,Schinnerer_D}. The Large project covered the entire $2\hspace{0.1cm}\mathrm{deg}^2$ COSMOS field, while the Deep covered only the central $\sim 1\hspace{0.1cm}\mathrm{deg}^2$. We use 1.4 GHz Large continuum mosaic \citep[hereafter 1.4 GHz Large map,][]{Schinnerer_L} characterized by $1.5^{\prime\prime}\times 1.4^{\prime\prime}$ resolution and $\approx 10.5 (15)\hspace{0.1cm}\mathrm{\mu Jy/beam}$ rms noise in the central 1(2) $\mathrm{deg}^2$. We also use 1.4 GHz Deep continuum mosaic \citep[hereafter 1.4 GHz Deep map,][]{Schinnerer_D}, produced from a combination of the earlier imaging data from the Large and later data from the Deep project. The final combined 1.4 GHz Deep map reaches a resolution of $2.5^{\prime\prime}$, and an average rms noise of $\approx 12\hspace{0.1cm}\mathrm{\mu Jy/beam}$. We construct the corresponding rms noise maps using AIPS RMSD 100 pixels circular mesh. We additionally use a joint radio source catalog produced from VLA-COSMOS Large and Deep catalogs as described in \citet{Schinnerer_D}.

We additionally use the MeerKAT telescope early observations. Under the MIGHTEE galaxy evolution survey ~\citep{Jarvis2016} $1.6\hspace{0.1cm}\mathrm{deg}^2$ of the COSMOS field was observed in L-band, centered at 1.284 GHz, with a single deep pointing. The total of 25 hours of observation resulted in the Early Science continuum data release \citep{Heywood2022}. The resulting radio continuum map exists in two variations: low resolution high sensitivity (reaching a resolution of $8.6^{\prime\prime}$ and rms of $2.7\hspace{0.1cm}\mathrm{\mu Jy/beam}$) and high resolution low sensitivity ($5^{\prime\prime}$ and $5.5\hspace{0.1cm}\mathrm{\mu Jy/beam}$). Using the former, a corresponding level 1 component catalog was produced as a part of the Early Science data release \citep{Heywood2022}. In this work, we use the low resolution high sensitivity map, the associated rms noise map provided by the authors, and the component catalog. As the map has a spatially varying effective frequency, we use the corresponding effective frequency map of \citet{Heywood2022} to find the frequency of observation at our sources' location (1.35 GHz), and update the cutout headers accordingly to ensure the correct frequency is used in the analysis.

At lower radio frequencies we use 325 MHz and 610 MHz radio continuum maps (hereafter 325 and 610 maps) generated from 45 h and 86 h of the Giant Meter Radio Telescope (GMRT) observations, respectively. The data reduction, imaging, cataloging and testing were done by \citet{Tisanic2019}. The 325 MHz map has a resolution of $10.8^{\prime\prime}\times 9.5^{\prime\prime}$ and median rms noise of $\approx 97\hspace{0.1cm}\mathrm{\mu Jy/beam}$, while 610 MHz map reaches a resolution of $5.6^{\prime\prime}\times 3.9^{\prime\prime}$, and a median rms of $\approx 39\hspace{0.1cm}\mathrm{\mu Jy/beam}$. We also use the corresponding rms noise maps and radio source catalogs provided by the authors.
\vspace{-0.2cm}
\section{Tailed radio morphology}
\label{sec:tailed_radio_morph}
\subsection{Wide-angle tailed radio galaxy 10913}
\label{sec:tailed_radio_morph_10913}
\begin{figure*}[]
    \centering
    \includegraphics[width=0.8\textwidth]{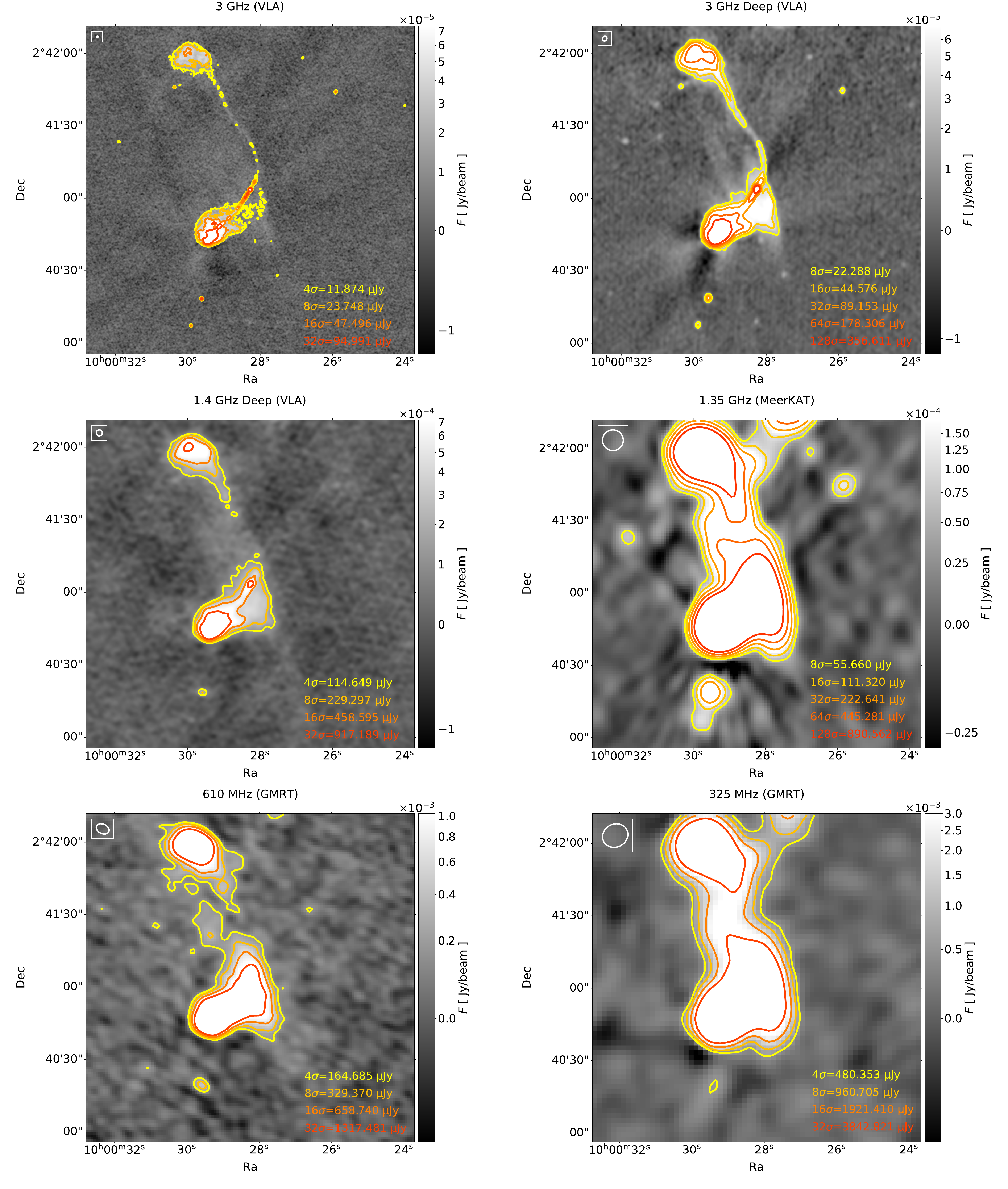}
    \caption{Radio images of wide angle tail radio galaxy 10913 - cutouts from 6 different radio maps/mosaics (see Section \ref{sec:data} for details and data references): 3 GHz, 3 GHz Deep, and 1.4 GHz Deep (VLA); 1.35 GHz (MeerKAT); 610 MHz and 325 MHz (GMRT), as indicated at the top of each panel. Radio contours are overlaid, with contour levels given in the bottom right; the corresponding radio beam is shown in the upper left. Contours are drawn at $n\sigma$, where $\sigma$ is the average local rms noise, estimated from the respective rms map cutout, and $n$ follows logarithmic ($\mathrm{log_{2}}$) scale. Images are displayed using an asinh scale from $-4\hspace{0.05cm}\sigma$ to $25\hspace{0.05cm}\sigma$.}
    \label{fig:fluxandcont10913}
\end{figure*}

Radio galaxy 10913 ($\mathrm{RA}=10^{\mathrm{h}}\hspace{0.02cm}0^{\mathrm{m}}\hspace{0.02cm}28.3^{\mathrm{s}}$, $\mathrm{DEC}=+02^{\circ}\hspace{0.02cm}41^{\prime}\hspace{0.02cm}3.48^{\prime\prime}$) is a WAT galaxy, hosted by the group's BGG - a giant, red, elliptical galaxy with an extended halo, as found in \citet{Vulic2025}. It is a typical FRII source \citep{FR1974}, launching two collimated radio jets ending in bright radio lobes. The jets are bent, most probably due to interaction (movement) with (through) the IGM at high relative velocity $\gtrsim 540\hspace{0.1cm}\mathrm{km/s}$ \citep{Vulic2025}. Here, in Fig.~\ref{fig:fluxandcont10913}, we present six cutouts from the radio maps at frequencies between 3 GHz and 325 MHz (data references in Section~\ref{sec:data}), in which the source is displayed together with the corresponding radio contours (details in figure caption).\par
For the analysis in this paper, we divide the galaxy's extended radio structure in six components: core, northern and southern jet, northern and southern lobe and south-west extended feature (SWEF). The components are best distinguishable at 3 GHz \citep[$0.75^{\prime\prime}$ resolution imaging data of][matching $\sim3.7$ kpc at at the group's redshift]{3ghz_smol_a}, as presented visually and labeled in Appendix~\ref{app:components_of_tailed_rgs} in the upper panel of Fig.~\ref{fig:plotcomp}. The radio galaxy in total extends over $\sim 90^{\prime\prime}$ in the north-south direction (measured at 3 GHz between the lobe tips in the plane of the sky), corresponding to $\sim 0.45\hspace{0.1cm}\mathrm{Mpc}$.
Both the northern and southern radio jets appear to originate directly from the core (i.e. within $\sim3.7$ kpc from it, set by our best resolution), and their projected lengths are $\sim320$ and  $\sim135$ kpc, respectively. The observed asymmetry in jet extent is most likely due to projection effects. Both jets end in bright radio lobes, the northern lobe $\sim90$ kpc wide (east-west direction), and $\sim65$ kpc long (north-south), and the southern lobe $\sim95$ kpc wide and $\sim80$ kpc long. We also identify an extended, irregular radio feature southwest of the core (SWEF), $\sim60$ kpc wide and $\sim95$ kpc long.\par
At some frequencies, galaxy components cannot be clearly separated due to limited resolution (and/or projection effects). This can introduce flux density mixing between neighboring components and may bias their relative contributions. To address this and/or acknowledge the limitations it imposes on the integrated flux density analysis (Section~\ref{sec:integrated_flux_density_spectra}), we apply the methods described in Appendix~\ref{app:components_of_tailed_rgs}.

\vspace{-0.2cm}
\subsection{Tailed radio galaxy 44}
\label{sec:tailed_radio_morph_44}
\begin{figure*}[]
    \centering
    \includegraphics[width=0.8\textwidth]{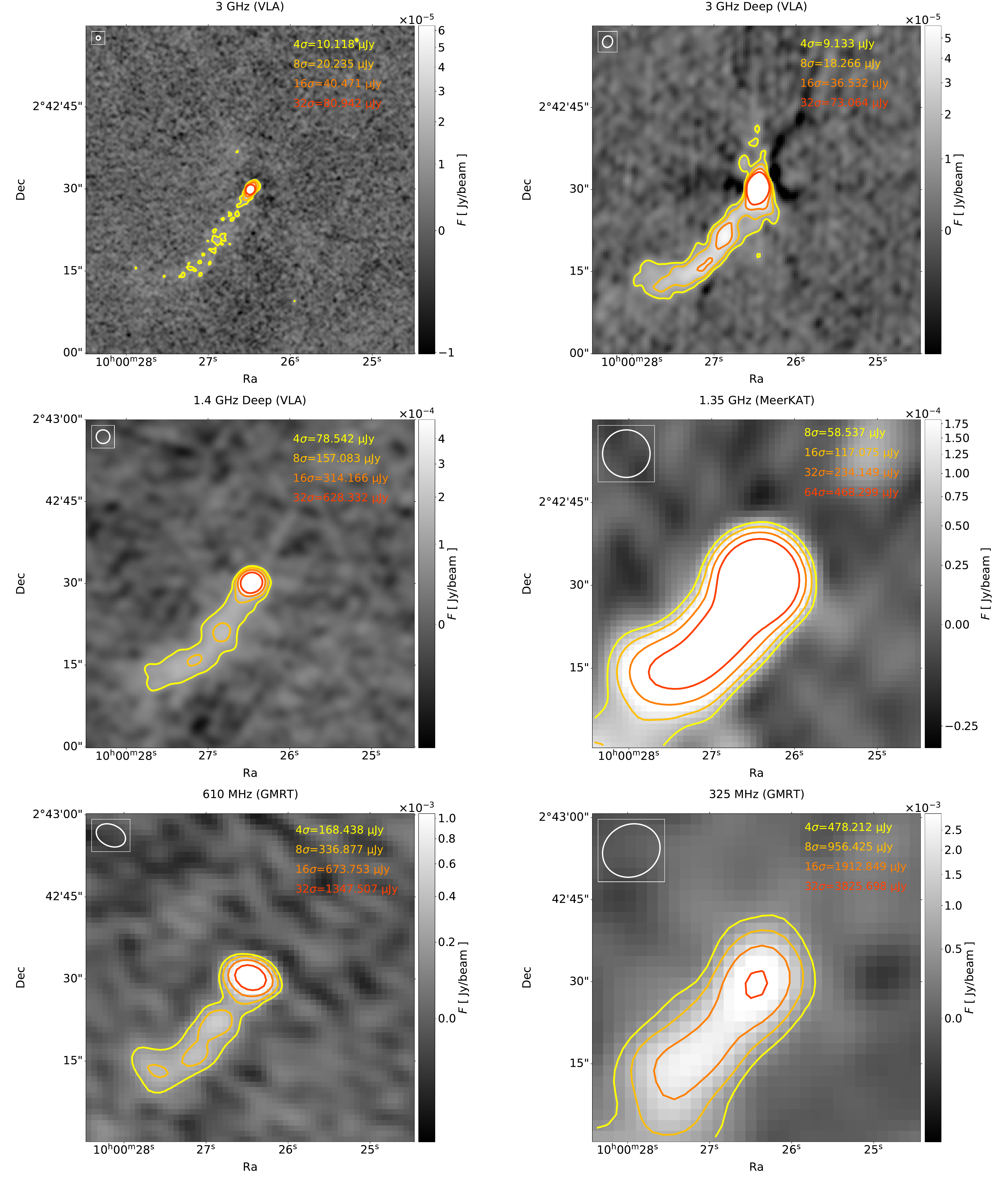}
    \caption{Radio images of tailed radio galaxy 44. Shown are cutouts from 6 radio maps/mosaics (see Section \ref{sec:data} for details and data references): 3 GHz, 3 GHz Deep, and 1.4 GHz Deep (VLA); 1.35 GHz (MeerKAT); 610 MHz and 325 MHz (GMRT), as indicated at the top of each panel. Radio contours are overlaid, with contour levels (see caption of Fig.~\ref{fig:fluxandcont10913} for details) indicated in the top right, and the corresponding radio beam in the upper left of each panel. Images are displayed using an asinh scale from $-4\hspace{0.05cm}\sigma$ to $25\hspace{0.05cm}\sigma$.}
    \label{fig:fluxandcont44}%
\end{figure*}

Radio galaxy 44 ($\mathrm{RA}=10^{\mathrm{h}}\hspace{0.02cm}0^{\mathrm{m}}\hspace{0.02cm}26.5^{\mathrm{s}}$, $\mathrm{DEC} = +02^{\circ}\hspace{0.02cm}42^{\prime}\hspace{0.02cm}29.88^{\prime\prime}$) is an FRI source \citep{FR1974}, most probably a HT galaxy, where jets stay unresolved even at the best resolution available to us \citep[$\sim3.7$ kpc resolution, 3 GHz VLA data of][see above]{3ghz_smol_a}. As discussed in \citet{Vulic2025}, this extreme jet bending could be caused by a large plane of the sky velocity of its host, or fast bulk IGM motion at the galaxy's location. This radio galaxy is hosted by the second brightest group galaxy, a giant, red, elliptical galaxy with extended halo, and somewhat less massive than the BGG (for details see \citealt{Vulic2025}). Fig.~\ref{fig:fluxandcont44} shows radio galaxy 44 (cutouts from maps) at different radio frequencies from 3 GHz to 325 MHz, and the corresponding radio contours (see caption for details).\par
In this paper, we divide radio galaxy 44 in two components: core and tail, as presented in the bottom panel of Fig.~\ref{fig:plotcomp}. At 3 GHz, the projected length of this radio galaxy is $\sim135$ kpc, measured between the core in the north-west and the tail tip in the south-east. For the first $\sim100$ kpc the tail is straight, and then slightly deviates (bends) northward (see Fig.~\ref{fig:fluxandcont44}).\par
Similar to radio galaxy 10913 (see the end of Sect.~\ref{sec:tailed_radio_morph_10913}), the two components here are not clearly separable at all frequencies due to resolution limits. For details on how this is handled in the analysis see Appendix~\ref{app:components_of_tailed_rgs} and Sect.~\ref{sec:integrated_flux_density_spectra}.
\vspace{-0.3cm}
\section{Integrated flux densities, luminosities and spectral indices}
\label{sec:integrated_flux_density_spectra}
\subsection{Objectives of the analysis}
\label{sec:IFDS_objectives}
To investigate their radio spectral evolution, we find integrated flux densities of tailed radio galaxies 10913 and 44 and of their individual components (as introduced in Section~\ref{sec:tailed_radio_morph}, also see Appendix~\ref{app:components_of_tailed_rgs}) at five frequencies: 3 GHz \citep{3ghz_smol_a}, 1.4 GHz \citep[Deep,][]{Schinnerer_D}, 1.35 GHz \citep{Heywood2022}, 610 MHz, and 325 MHz \citep{Tisanic2019}.
We use these, consistently derived integrated flux densities of radio galaxies rather than the catalog values (for details on radio source catalogs see Section~\ref{sec:data} and references therein). We do so to avoid any possible difficulties and limitations in the analysis due to differences in the methods used to generate catalog values, and/or the lack of reported values in different surveys. 
To study the radio spectral evolution of the galaxies and their components, we plot integrated flux densities against frequency on a $\ln-\ln$ scale to derive their spectral indices. We also calculate the corresponding radio luminosities.
\vspace{-0.4cm}
\subsection{Method}
\label{sec:IFDS_method}

We find integrated flux densities of radio galaxies 10913 and 44, at each of the five radio frequencies, by integrating over pixels in the corresponding radio maps with flux density above $3\sigma$, and dividing the sum with the corresponding beam size in pixels. Here, $\sigma$ is the local rms noise for a given source at a given frequency (estimated as described and used in contour plotting in Section~\ref{sec:tailed_radio_morph}). To estimate uncertainties, we also apply this at $2\sigma$ and $4\sigma$ thresholds, and use deviations from the $3\sigma$ value as the upper and lower errors on the integrated flux density, respectively. We previously employed this error estimation method in \citet{Vulic2025}, where we discuss its advantages and limitations.\par
To estimate integrated flux densities of the individual components (Section~\ref{sec:tailed_radio_morph}) of the two radio galaxies (and their errors), we apply the above described method, only integrating within the corresponding component regions (see Appendix~\ref{app:components_of_tailed_rgs}).
We only calculate this for components classified as separable at a given frequency (see Section~\ref{sec:defining_and_separating_galaxy_comp} for the definition and details), i.e. when the flux density mixing between the neighboring components is not significantly affecting the accuracy of the results.
In both galaxies, we were only able to estimate the core flux density at 3 GHz \citep[data of][]{3ghz_smol_a}, and at 1.4 GHz \citep[Deep,][]{Schinnerer_D}, as the peak value within the core region resulting from the Gaussian fit (see Section~\ref{sec:defining_and_separating_galaxy_comp}). The corresponding error was set to $1\sigma$ - the local rms noise. At lower frequencies (1.35 GHz, 610 MHz, 325 MHz) we do not attempt to find core flux densities, since at these resolutions the jets seem to interfere with the core flux density at the central brightest pixel in the core region.\par
We plot the integrated flux densities of the galaxies and of their individual components against frequency on a $\ln-\ln$ scale, and find their respective spectral indices by applying the least squares fit to the obtained data points (assuming a linear model), together with the associated errors, quantifying the scattering around the best-fit line. When only two data points were available, asymmetric upper and lower slope uncertainties were estimated as the difference between the best-fit slope and the steepest and flattest slope allowed by the data, within their respective error ranges, respectively.\par
We derive monochromatic radio luminosities $L_{\nu}$ of the two tailed radio galaxies, 10913 and 44, at five frequencies: 3 GHz \citep{3ghz_smol_a}, 1.4 GHz \citep[Deep,][]{Schinnerer_D}, 1.35 GHz \citep{Heywood2022}, 610 MHz, and 325 MHz \citep{Tisanic2019}, in the source rest-frame. We do so using the following expression: $L_{\nu}=4\pi F_{\nu} D_L^2/(1+z)^{1-\alpha}$. Here, $F_{\nu}$ and $\alpha$ are the radio galaxy's flux density at frequency $\nu$ and galaxy's spectral index, respectively, both found as described above. $D_L$ is the luminosity distance, and $z$ is the spectroscopic redshift.
\vspace{-0.3cm}
\subsection{Results}
\label{sec:IFDS_results}
The resulting integrated flux densities of the two tailed radio galaxies at five radio frequencies (from 325 MHz to 3 GHz) are listed in the upper part of Table~\ref{table:fluxlum}.
We compare them with values from the corresponding radio source catalogs when the latter is available, as presented visually in Appendix~\ref{app:comparison_fluxes}. We find the values agree well with the relative difference (originating from differences in methods) below $10\%$ on average, i.e. within the reported error ranges (see Fig.~\ref{fig:comparison_of_fluxes_with_catalog}), and the largest relative difference $\approx 18\%$ for radio galaxy 44 at 3 GHz \citep[data of][]{3ghz_smol_a}. The latter arises because much of the tail’s 3 GHz flux density is between our $3\sigma$ threshold and the $5\sigma$ limit used by \citet{3ghz_smol_a}.\par
\begin{figure*}[h!]
    \centering
    \includegraphics[width=\hsize]{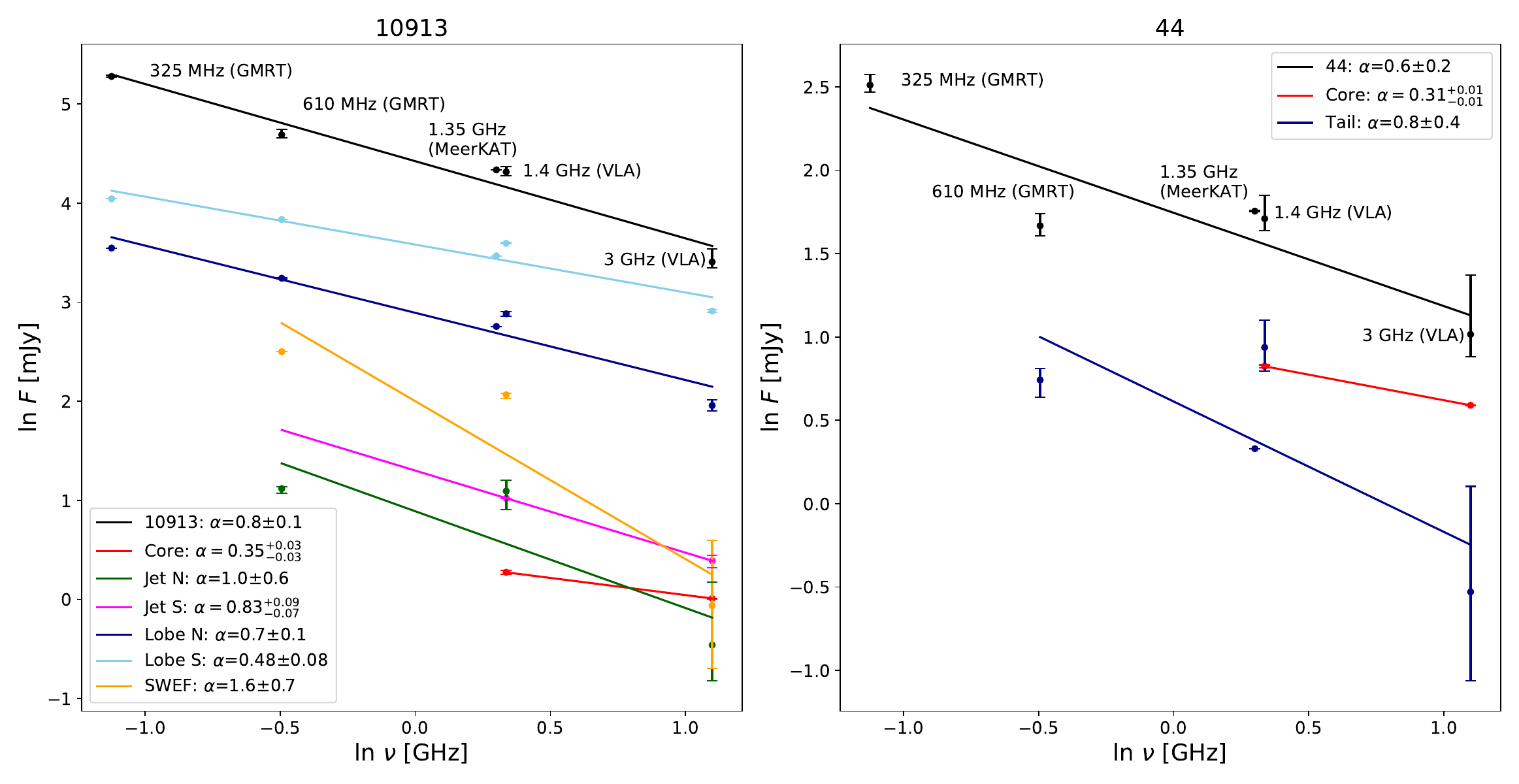}
    \caption{Integrated flux density versus frequency (radio spectra) in $\ln$ space for radio galaxies (and their individual components): 10913 (left) and 44 (right). To obtain the data points (i.e. integrated flux densities; see Section~\ref{sec:integrated_flux_density_spectra}) and their uncertainties, we used radio maps at five frequencies: 3 GHz (VLA), 1.4 GHz Deep (VLA), 1.35 GHz (MeerKAT), 610 MHz (GMRT), and 325 MHz (GMRT). The straight lines represent the best-fit linear models, with slopes corresponding to the derived spectral indices (see legends).}
    \label{fig:ln_flux_ln_freq}%
\end{figure*}

The galaxies' integrated flux densities and those of their individual components, plotted against frequency, are shown in the left and right panels of Fig.~\ref{fig:ln_flux_ln_freq} for radio galaxies 10913 and 44, respectively. Spectral indices were derived from these plots (see Section~\ref{sec:IFDS_method}), $\alpha=0.8\pm 0.1$ for 10913 and $\alpha=0.6\pm 0.2$ for radio galaxy 44. In radio galaxy 10913, the core has the flattest spectrum with $\alpha=0.35\pm0.03$. Both jets and lobes have steeper spectra than the core. For the southern jet we find $\alpha=0.83\substack{+0.09 \\ -0.07}$, consistent (within error ranges) with the value we obtain for the northern jet ($\alpha=1.0\pm 0.6$). Furthermore, the spectral indices of the northern ($\alpha=0.7\pm 0.1$) and southern ($\alpha=0.48\pm 0.08$) lobes are flatter than those of the jets. For the SWEF (see Section~\ref{sec:tailed_radio_morph} and Appendix~\ref{app:components_of_tailed_rgs}) we find $\alpha=1.6\pm 0.7$, the steepest value among the components of radio galaxy 10913. In case of radio galaxy 44, we find $\alpha=0.31\pm 0.01$ for the core, which is flatter than the tail with $\alpha=0.8\pm0.4$.\par
The resulting radio luminosities are presented in the bottom part of Table~\ref{table:fluxlum}. Both galaxies display consistently high radio luminosities across all five frequencies. For the WAT radio galaxy 10913 luminosities range from $1.19\substack{+0.16 \\ -0.08}\times 10^{25}\hspace{0.1cm}\mathrm{W}\hspace{0.05cm}\mathrm{Hz^{-1}}$ at 3 GHz to $7.7\substack{+0.2 \\ -0.2}\times 10^{25}\hspace{0.1cm}\mathrm{W}\hspace{0.05cm}\mathrm{Hz^{-1}}$ at 325 MHz, while for the tailed radio galaxy 44 we find $1.0\substack{+0.4 \\ -0.1}\times 10^{24}\hspace{0.1cm}\mathrm{W}\hspace{0.05cm}\mathrm{Hz^{-1}}$ at 3 GHz and $4.6\substack{+0.4 \\ -0.3}\times 10^{24}\hspace{0.1cm}\mathrm{W}\hspace{0.05cm}\mathrm{Hz^{-1}}$ at 325 MHz.

\begin{table*}
\caption{Monochromatic integrated flux densities and rest-frame luminosities at five frequencies: 3 GHz (VLA), 1.4 GHz Deep (VLA), 1.35 GHz (MeerKAT), 610 MHz (GMRT), and 325 MHz (GMRT), for tailed radio galaxies 10913 and 44.}
\label{table:fluxlum}      
\centering          
\begin{tabular}{c | c c c c c  }    
\hline\hline       
 & \multicolumn{5}{c}{$F_{\nu}\hspace{0.2cm}[\mathrm{mJy}]$}\\ \hline
   Source      & $3\hspace{0.1cm}\mathrm{GHz}$ & $1.4\hspace{0.1cm}\mathrm{GHz}$ & $1.35\hspace{0.1cm}\mathrm{GHz}$ & $610\hspace{0.1cm}\mathrm{MHz}$ & $325\hspace{0.1cm}\mathrm{MHz}$ \\
\hline                    
   10913 & $30\substack{+4\\-2}$ & $75\substack{+4 \\ -3}$ & $76.3\substack{+0.2 \\ -0.1}$ & $109\substack{+6 \\ -3}$ & $196\substack{+3 \\ -1}$\\
   44 & $2.8\substack{+1.0 \\ -0.4}$ & $5.5\substack{+0.8 \\ -0.4}$ & $5.78\substack{+0.02 \\ -0.02}$ & $5.3\substack{+0.4 \\ -0.3}$ & $12.3\substack{+0.8 \\ -0.5}$\\ \hline \hline
  &\multicolumn{5}{c}{$L_{\nu}\hspace{0.2cm}[10^{24}\hspace{0.1cm}\mathrm{W\hspace{0.05cm}Hz^{-1}}]$}\\ \hline
 Source & $3\hspace{0.1cm}\mathrm{GHz}$ & $1.4\hspace{0.1cm}\mathrm{GHz}$ & $1.35\hspace{0.1cm}\mathrm{GHz}$ & $610\hspace{0.1cm}\mathrm{MHz}$ & $325\hspace{0.1cm}\mathrm{MHz}$ \\     
  \hline
 10913 & $11.9\substack{+1.6 \\ -0.8}$ & $29\substack{+2 \\ -1}$ & $30.1\substack{+0.8 \\ -0.8}$ & $43\substack{+3 \\ -2}$ & $77\substack{+2 \\ -2}$\\ 
  44 &  $1.0\substack{+0.4 \\ -0.1}$ & $2.1\substack{+0.3 \\ -0.2}$ & $2.2\substack{+0.1 \\ -0.1}$ & $2.0\substack{+0.2 \\ -0.1}$ & $4.6\substack{+0.4 \\ -0.3}$\\ \hline \hline
\end{tabular}
\end{table*}
\vspace{-0.3cm}
\section{Spatially resolved spectral characteristics}
\label{sec:specindmaps}
\subsection{Objectives and method}
To investigate the spatial distribution of the spectral index within the radio galaxies we create pixel-by-pixel spectral index maps and the corresponding error maps. We generate the maps using the Broadband Radio Astronomy Tools (BRATS) software \citep{harwood2013,harwood_hardcastle_croston2015}, and radio maps at 3 GHz (VLA), 1.4 GHz Deep (VLA) and 610 MHz (GMRT).\par
Images (cutouts) at different frequencies were first smoothed and regridded to match the resolution of the 610 MHz map using CASA tasks \texttt{imsmooth} and \texttt{imregrid} \citep{casa}. We checked in DS9 \citep{DS9} whether the integrated flux density within the region surrounding each galaxy is the same before and after the CASA processing.
Values agree within less than 0.5\% implying the processing was done properly and kept the physical information unchanged. For each source, the images were already well coinciding in astrometry, however, we checked for any possible (minor) misalignment by fitting a 2D Gaussian to a nearby high signal-to-noise ratio (S/N) point-like source (for radio galaxy 10913) or to the galaxy core itself (for radio galaxy 44, in the lack of an adequate high S/N point-like source in its vicinity). Any differences in the center coordinates of a 2D Gaussian found at 3 GHz and 1.4 GHz compared to 610 MHz were corrected for in the corresponding image headers. Proper alignment is crucial to avoid artifacts in the analysis at the edges of radio galaxies. We repeated the 2D Gaussian fitting on the corrected images to verify the alignment. For radio galaxy 10913, the initial offsets relative to the 610 MHz map were $\approx 1.2^{\prime\prime}$ and $\approx0.55^{\prime\prime}$, for the 3 GHz and 1.4 GHz Deep data, respectively. For radio galaxy 44, the corresponding offsets were $\approx 0.65^{\prime\prime}$ and $\approx0.3^{\prime\prime}$. In the second fitting, the misalignment was reduced to $\lesssim0.1^{\prime\prime}$ in all cases.\par

Along with the radio images, we load into BRATS two region files: a rectangular area enclosing the source where the calculations are performed, and a circular region within the image, containing only background noise, which BRATS uses to estimate the local rms noise $\sigma$. BRATS $\sigma$ values are consistent with those we estimated and used for contour plotting (Section~\ref{sec:tailed_radio_morph}) and integrated flux density analysis (Section~\ref{sec:integrated_flux_density_spectra}). After loading the images, we only keep the pixels above $5\sigma$ (by setting \texttt{sigma}=5). We set flux calibration errors to 5\% for 3 GHz and 1.4 GHz Deep, and to 10\% for 610 MHz data. We set the on-source multiplier to 3 and S/N expected from a region to 1 (default value). Then, with \texttt{setregions} we divide the source into pixel-size regions where we perform spectral index calculation. The calculation itself is done with \texttt{specindex} command in BRATS.
For three frequencies, by default, BRATS calculates spectral indices using a GNU scientific libraray (GSL)\footnote{\url{https://www.gnu.org/software/gsl/}} weighted least-squares fit, with weights $w=1/\sigma^2$, where $\sigma$ is the error on a given flux density measurement (for details on this and the associated uncertainty estimation in BRATS, see the BRATS cookbook\footnote{\label{bratsfootnote}\url{https://github.com/JeremyHarwood/BRATS}}). In the final spectral index (and error) maps we manually remove a small number of pixels, corresponding to very high, nonphysical spectral indices, located on the galaxies' edges.
\vspace{-0.2cm}
\subsection{Results}
The resulting spectral index maps (and the corresponding error maps) are presented along with the results of spectral age analysis (see Section~\ref{sec:spectral_age_analysis}) in the top left (and right) panels in Figs.~\ref{Fig:BRATS_specage_10913} and ~\ref{Fig:BRATS_specage_44}, for radio galaxies 10913 and 44, respectively. We overlay 610 MHz radio contours on top of the presented maps.\par
In radio galaxy 10913, the spectral index gradually steepens along the northern jet and flattens in the northern lobe, particularly near its outer edge. In the south, at the 610 MHz resolution used in this analysis (and likely also due to projection effects; see Sections~\ref{sec:tailed_radio_morph} and \ref{sec:defining_and_separating_galaxy_comp}), the flux densities of the southern jet and lobe blend, with the latter dominating. This produces the gradual spectral flattening observed from the core toward the southern lobe tip. Flattening in the lobes is typical of FRII sources and is likely related to electron re-acceleration. The steepest spectral index occurs southwest of the core, corresponding to the SWEF component. In radio galaxy 44, the spectral index gradually steepens along the tail, from the core toward the tail tip. However, a localized flattening occurs along the jet (at $\mathrm{RA}\approx10^{\mathrm{h}}\hspace{0.02cm}0^{\mathrm{m}}\hspace{0.02cm}27^{\mathrm{s}}$, $\mathrm{DEC}\approx+02^{\circ}\hspace{0.02cm}42^{\prime}\hspace{0.02cm}20^{\prime\prime}$), possibly indicating a hotspot (i.e., a place of electron re-acceleration), uncommon for FRI sources such as this galaxy. Note, however, that the uncertainties are also high in this case (top right panel in Fig.~\ref{Fig:BRATS_specage_44}).\par

We repeated the spectral index analysis combining the 3 GHz Deep (VLA) and 1.4 GHz Large (VLA) maps (for details on data see Section~\ref{sec:data}). This yielded higher resolution but two-frequency (rather than three-frequency based) maps. The results, presented in Appendix~\ref{app:two-frequency-based-specindmaps}, follow the trends in the distribution of the spectral index found here. However, owing to the improved resolution, and combined with the high resolution 3 GHz radio images (Section~\ref{sec:tailed_radio_morph}), we identify five localized spot-like regions of enhanced radio brightness along the northern jet of the WAT radio galaxy 10913. These regions show associated local spectral index flattening, which is clearly detected for the brighter features and only indicated for the fainter ones due to limited S/N. No known background radio sources are present at the locations of these features. Their physical interpretation is discussed in Section~\ref{sec:discussion}.
Further improvements in the analysis would require better resolution data at MHz frequencies and observations at higher radio frequencies (above 3 GHz) in this part of the COSMOS field.
\vspace{-0.3cm}
\section{Radio galaxies' age estimates}
\label{sec:age_analysis}
\subsection{Spectral age analysis}
\label{sec:spectral_age_analysis}
\subsubsection{Aging models and BRATS implementation}
\label{sec:specagetheory}
The extended structure of radio galaxies is shaped by a combination of internal dynamical processes, such as jet launching and propagation, and by interactions with the surrounding IGM or ICM. Here, we employ spectral aging analysis to trace the history of radiative losses of the electrons in the ejected plasma, which approximately reflect the underlying dynamical evolution. In FR II sources \citep{FR1974} relativistic jets of magnetized plasma are launched from the central AGN and can extend up to Mpc scales (see \citealt{saikia2022} for a recent overview of observational jet properties). The jets terminate in hotspots, where they encounter the ICM or IGM, resulting in a shock that strongly accelerates plasma electrons. As the hotspots advance, they leave behind accelerated plasma in the form of lobes, which age through synchrotron and inverse-Compton losses. Assuming a uniform magnetic field within the emitting region, higher-frequency electrons radiate energy away more rapidly, namely, they age (cool) faster. This produces a progressive steepening of the emission spectrum (i.e., the spectral index) toward higher frequencies. Two widely used spectral aging models that assume a constant magnetic field strength $B$ are Kardashev and Pacholczyk (KP) model \citep{kardashev1962, pacholczyk1970} and Jaffe and Perola (JP) model \citep{jaffeperola1973}.  A more complex alternative is the Tribble model (TRIBJP, following the BRATS abbreviation scheme), which allows for spatial variations of $B$ within the source \citep{Tribble1993}. For a detailed theoretical description, we refer the reader to the original works.\par
Here, we use BRATS software \citep{harwood2013,harwood_hardcastle_croston2015} to perform the spectral age analysis. BRATS fits the above spectral aging models to the observed broad-band data on a region-to-region (in our case pixel-to-pixel) basis. For a given model family (JP, KP or TRIBJP), it generates multiple models with different spectral ages drawn from a user defined range [$t_{min}$,$t_{max}$], and calculates the corresponding model flux densities $S_{model,\nu}$, at each frequency $\nu$ included in the analysis.
The computation of model flux densities in BRATS requires user-defined parameters such as the magnetic field strength $B$ (see the next Section) and the injection index $\alpha_{inj}$ (related to the power-law exponent of the initial electron energy distribution $\delta=2\alpha_{inj}+1$). At each pixel, BRATS evaluates the goodness of fit for each particular model by finding the $\chi^2$ value,
$
\chi^2=\sum_{\nu=1}^{N} \bigg( \frac{S_{i,\nu}-S_{model,\nu}}{\Delta S_{i,\nu}} \bigg )^2
.$
Here, $S_{i,\nu}$ is the observed flux density at the $i$-th pixel at frequency $\nu$, $S_{model,\nu}$ the model flux density at frequency $\nu$, and $\Delta S_{i,\nu}$ is the total uncertainty for the $i$-th pixel (see BRATS cookbook\footnote{See footnote \ref{bratsfootnote}} for details). At each pixel, BRATS compares the goodness of fit among different models, identifying the best-fit (best $\chi^2$) model, that is, the spectral age. The corresponding error is found as $1\sigma$ deviation from the minimum of the $\chi^2$ distribution. The resulting spectral ages and their uncertainties can be visualized as spectral age and error maps. For model implementation details in BRATS refer to \citet{harwood2013, harwood_hardcastle_croston2015}.
\vspace{-0.2cm}
\subsubsection{Data and parameters of the analysis}
\label{sec:dataandparametersforthespectralageanalysis}
We perform the spectral age analysis with BRATS for radio galaxies 10913 and 44, fitting JP, KP, and TRIBJP models, and using broadband radio data (maps) at 3 GHz (VLA), 1.4 GHz Deep (VLA) and 610 MHz (GMRT). The galaxy images at different frequencies were smoothed, regridded to a common resolution (that of the 610 MHz data), and astrometrically aligned as previously described in Section~\ref{sec:specindmaps}, where the same datasets were used to generate the spectral index maps. We excluded the AGN core from this analysis so that it would not interfere with the subsequent BRATS statistical analysis, as reasoned in the BRATS cookbook.

For each galaxy, we estimate the lobe magnetic field strength $B$, treated as fixed in the JP and KP models and as a mean value in the TRIBJP model (see \citealt{harwood2013} for details). We estimate $B$ using Python interface for synchrotron libraries PySYNCH\footnote{\url{https://github.com/mhardcastle/pysynch}}, namely, the \texttt{synchro} code \citep{Hardcastle1998}.
With \texttt{synchro} we compute the synchrotron emissivity assuming a user-defined power-law electron energy distribution, within the user-defined regions encompassing radio lobes. We assume a power-law spectrum with $\delta=2$ (i.e. $\alpha_{inj}=0.5$), and $\gamma$ from 1 to $10^{6}$, and approximate the lobes (i.e., the tail in case of radio galaxy 44), with ellipsoids at the corresponding galaxy's spectroscopic redshift. The magnetic field is then adjusted until the model emissivity matches (is normalized to) the observed flux density at 610 MHz (which is the lowest frequency included in the analysis and the least affected by aging). For radio galaxy 10913, we calculate $B$ in each lobe separately and then use the average in the spectral age analysis. Furthermore, in calculations we assume only electrons (i.e., there are no non-radiating particles), and equipartition of energy between the magnetic field and the relativistic particles. However, some studies show that the magnetic field can differ from the equipartition value for a median factor of $\approx0.4$ \citep{ineson2017}. Here, we perform the spectral age analysis assuming both cases, the equipartition value $B_{eq}$ and $0.4$ of $B_{eq}$. This resulted in analyzing 6 different cases (models) corresponding to the combinations of each magnetic field strength with each of the above three aging models. For the resulting spectral ages and the systematic offsets between the values derived under different magnetic field assumptions see Section~\ref{sec:resultsofthespectralageanalysis} and Appendix~\ref{app:specage_analysis_details}.

We set the injection index in BRATS for each source and model to $\alpha_{\rm{inj}}=0.5$. This value is consistent with previous work (e.g. \citealt{jaffeperola1973}; \citealt{carilli1991}; \citealt{orru2010}), who typically find $\alpha_{\rm{inj}}\sim0.5-0.7$, although steeper values up to 0.86 have been reported \citep{harwood2013}. The latter study also introduced the BRATS \texttt{findinject} tool. We used \texttt{findinject} here to search for the best-fit injection index for both sources and all assumed models (6 cases, see above), adopting $\alpha_{\rm{inj}}=0.5$ (theoretical lower limit; \citealt{bell1978}) as the lower and $\alpha_{\rm{inj}}=1$ as the upper bound. In all cases, the best-fit value converges to 0.5. We note that should the true injection index be steeper than 0.5, the resulting spectral ages would be systematically lower.
We assumed $\gamma$ between 1 and $10^{6}$ in the spectral age analysis. We fixed the spectral age fitting range to $0\hspace{0.1cm}\rm{Myr} - 50\hspace{0.1cm}\rm{Myr}$. We chose the age resolution (step) of $2\hspace{0.1cm}\rm{Myr}$ (\texttt{ageres} $=26$) and adjusted the precision of fitting by setting \texttt{levels} to 5.

\begin{figure*}[h!]
    \centering
    \includegraphics[width=0.78\textwidth]{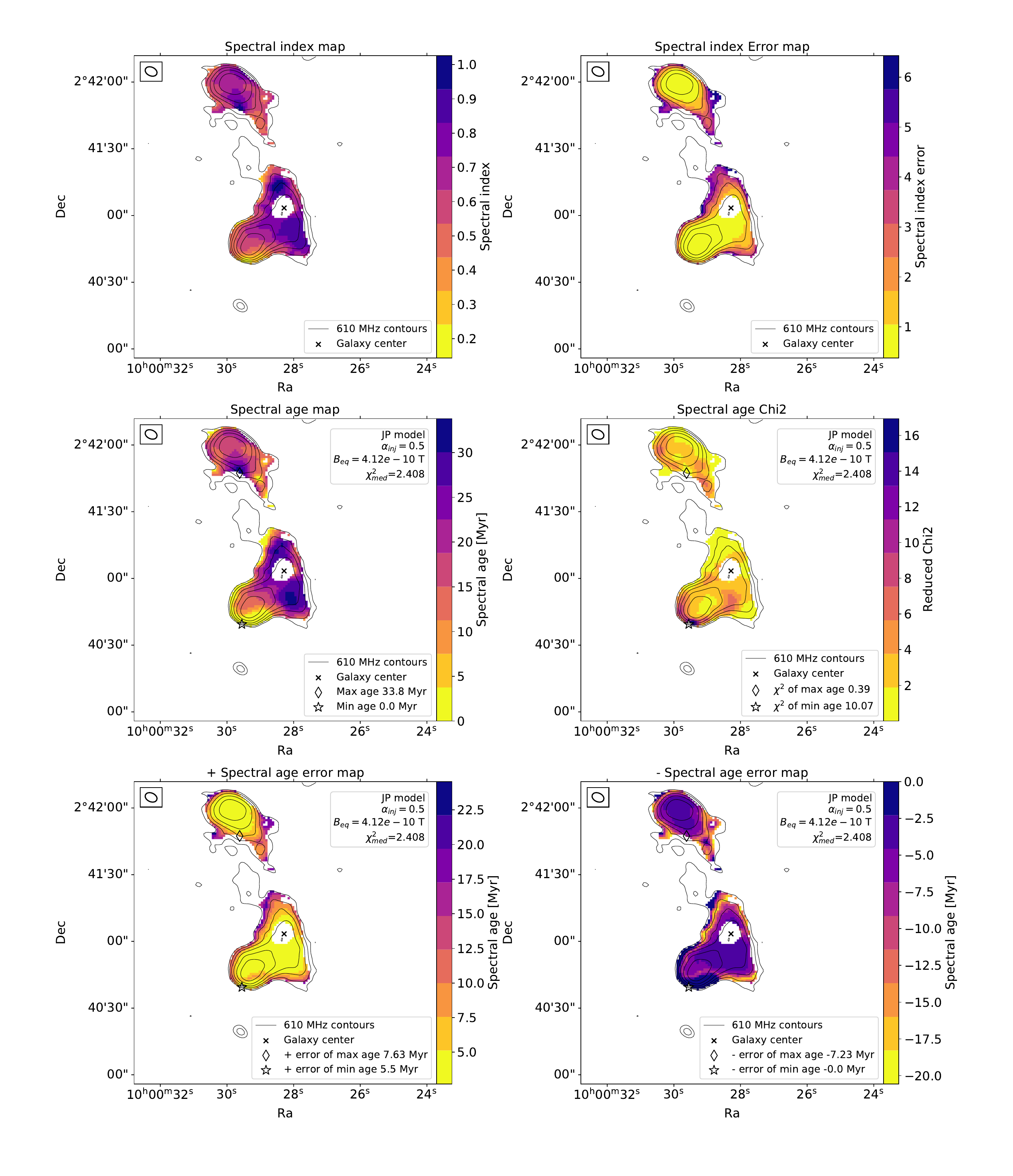}
    \caption{Results of BRATS analysis for radio galaxy 10913. The top two panes show a spectral index map and the corresponding error map (Section~\ref{sec:specindmaps}). The bottom four panels display the maps resulting from the spectral age analysis: a spectral age map, the corresponding $\chi^2$, and positive and negative error maps. Maps are presented for the best-fit model, JP, assuming $B_{eq}$. A diamond and a star mark the location of the oldest and the youngest spectral age, respectively. In all panels, we overlay the 610 MHz radio contours.}
    \label{Fig:BRATS_specage_10913}%
\end{figure*}
\begin{figure*}[h!]
    \centering
    \includegraphics[width=0.782\textwidth]{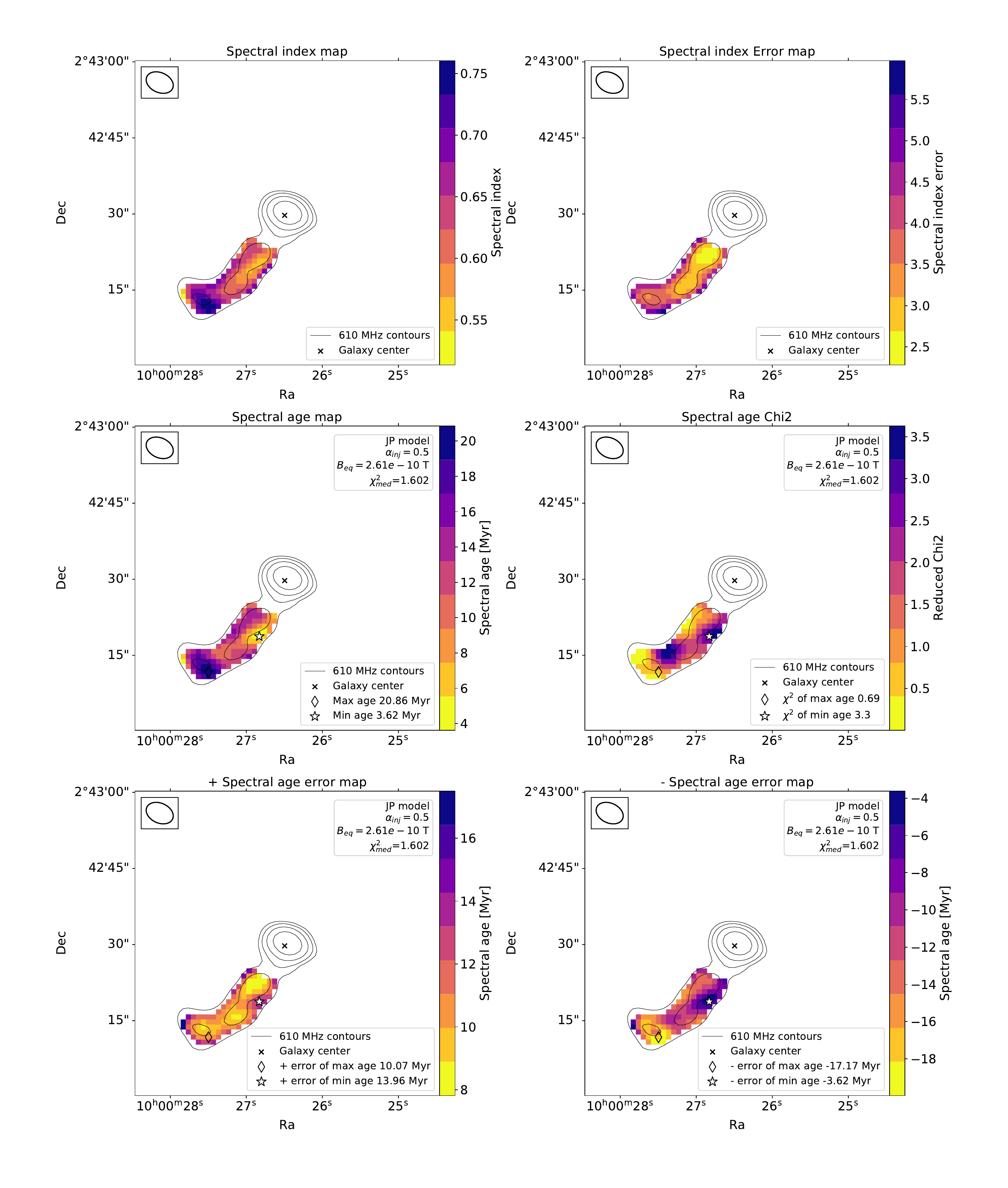}
    \caption{Results of BRATS analysis for radio galaxy 44. The panel layout in the figure and the labels are the same as in Fig.~\ref{Fig:BRATS_specage_10913}. Maps are presented for the best-fit model, JP, assuming $B_{eq}$.}
    \label{Fig:BRATS_specage_44}
\end{figure*}
\vspace{-0.3cm}
\subsubsection{Results}
\label{sec:resultsofthespectralageanalysis}
\begin{table}
\caption{The minimal and maximal ages resulting from BRATS spectral age analysis of radio galaxies 10913 and 44.}
\label{table:BRATSfinaloutput}      
\centering          
\begin{tabular}{c | c c c c  }    
\hline\hline  
ID    & Model & $\chi^2$ & $\tau_{min}$[Myr] & $\tau_{max}$[Myr] \\ \hline \\
10913 &   JP    &   $2.408$   &    $0.00\substack{+5.50 \\ -0.00}$     &     $33.80\substack{+7.63 \\ -7.23}$    \\ \\
44    &   JP    &   $1.602$   &    $3.62\substack{+13.96 \\ -3.62}$     &  $20.86\substack{+10.07 \\ -17.17}$   \\ \\
\hline
\end{tabular}
\tablefoot{
The minimal and maximal ages presented here are derived assuming the JP model \citep{jaffeperola1973} and the equipartition magnetic field $B_{eq}$. For details on spectral age analysis and results derived assuming different models (cases), see Section~\ref{sec:spectral_age_analysis} and Appendix~\ref{app:specage_analysis_details}.}
\end{table}
All 6 cases (models) for both galaxies converged at the 90\% significance level based on the median binning. We produced spectral age maps and the corresponding $\chi^2$, and positive and negative error maps. The resulting maps are only presented for the best-fit model (lowest median reduced $\chi^2$), which is, for both galaxies, the JP model assuming $B_{eq}$, and are shown in the middle and the bottom two panels of Figs.~\ref{Fig:BRATS_specage_10913} and \ref{Fig:BRATS_specage_44} for radio galaxies 10913 and 44, respectively. Also shown in the top two panels of these figures are the spectral index map and the corresponding error map (for details see Section~\ref{sec:specindmaps}). In all panels, we overlay the 610 MHz contours (corresponding to the ones from figures \ref{fig:fluxandcont10913} and \ref{fig:fluxandcont44}). For each galaxy, the minimum and maximum spectral ages from the best-fit model are listed in Table~\ref{table:BRATSfinaloutput}, with the corresponding pixels marked by a star and a diamond in Figs.~\ref{Fig:BRATS_specage_10913} and \ref{Fig:BRATS_specage_44}. The minimum and maximum ages of all other models are provided in Appendix~\ref{app:specage_analysis_details}, where results for all six cases are summarized and compared (Tables~\ref{table:BRATSallresults10913}, \ref{table:BRATSallresults44}; Figs.~\ref{Fig:modelcomp10913}, \ref{Fig:modelcomp44}).\par 
In brief, the TRIBJP model yields the highest maximum ages ($\tau_{\rm{max}}$), while the KP model gives the lowest. The difference between the two is $\approx 4-5\hspace{0.1cm}\rm{Myr}$ and $\approx 3-4\hspace{0.1cm}\rm{Myr}$ for 10913 and 44, respectively, and is largely independent of the assumed magnetic field strength. Between the two magnetic field assumptions, all three models (JP, KP, TRIBJP) show a systematic shift toward lower $\tau_{\rm{max}}$ in the non-equipartition case, by $\approx 4\hspace{0.1cm}\rm{Myr}$ for 10913 and $\approx 5.5\hspace{0.1cm}\rm{Myr}$ for 44.
However, all the above offsets fall within the corresponding error ranges, namely, given the sensitivity and resolution of our data, the different models yield mutually consistent spectral ages. We therefore disregard these differences and, from here on, discuss only the results of fitting the JP model assuming $B_{eq}$ for each galaxy. We caution that data constrain the model only up to 3 GHz.

For radio galaxy 10913, the analysis yielded a maximum age $\tau_{\rm{max}}=33.80\substack{+7.62 \\ -7.23}\hspace{0.1cm}\mathrm{Myr}$, with the oldest plasma located at the bottom of the northern lobe (diamond mark in Fig.~\ref{Fig:BRATS_specage_10913}). This likely corresponds to plasma ejected from the core that reached the lobe but was not re-accelerated and simply aged. We also identified very old plasma in the SWEF, possibly material that was ejected from the core, however, instead of reaching the lobes through a collimated jet, spilled sideways and continued to age. The youngest plasma is typically found in the core and hotspots; however, the core was masked in the analysis for the reasons outlined in Section~\ref{sec:dataandparametersforthespectralageanalysis}. The youngest 
detected regions therefore lie within the lobes. The model yielded a minimum age $\tau_{\rm{min}}=0.00\substack{+5.49 \\ -0.00}\hspace{0.1cm}\mathrm{Myr}$, located near the southern lobe edge (star mark in Fig.~\ref{Fig:BRATS_specage_10913}).We also detect young plasma in a hotspot along the northern jet, just before it merges into the lobe.

For radio galaxy 44, the spectral age analysis assuming the JP model and $B_{\rm{eq}}$ resulted in a maximum age $\tau_{\rm{max}}=20.86\substack{+10.07 \\ -17.17}\hspace{0.1cm}\rm{Myr}$ at the end of its tail. This oldest plasma likely originated in the core and traveled through the IGM in a non-collimated flow, forming the tail, where it subsequently aged without further re-acceleration. The spectral age map (in Fig.~\ref{Fig:BRATS_specage_44}) also shows a general increase in age with distance from the core, as expected for an FRI source. Given the latter and that the core was masked during fitting, the minimum detected age is greater than zero, $\tau_{\rm{min}}=3.62\substack{+13.96 \\ -3.62}\hspace{0.1cm}\mathrm{Myr}$. The low spectral ages found at the far eastern edge of the tail are most likely edge artifacts, consistent with their very large associated errors (see the positive error map in Fig.~\ref{Fig:BRATS_specage_44}). In general, the spectral age uncertainties for this source are noticeably larger than those for radio galaxy 10913, most likely due to resolution limits in the analysis. We further discuss the results in Sections~\ref{sec:prop_and_evol_jets} and ~\ref{sec:discussion_specdynage}.
\vspace{-0.3cm}
\subsection{Dynamical age analysis}
\label{sec:dyn_age_analysis}
\vspace{-0.1cm}
\subsubsection{Method}
We estimated the dynamical ages of radio galaxies 10913 and 44 following the approach of \citet{ineson2017}. Using the IGM temperature derived from X-ray spectra in \citet{Vulic2025}, we computed the sound speed in the IGM and adopted it as the expansion velocity:
\begin{equation}
    c_{IGM}=v_{exp}=\sqrt{\frac{\Gamma kT}{\mu m_H}}
\end{equation}
Here $\Gamma=5/3$ is the adiabatic factor for a monoatomic gas, $kT=2.4\pm0.6$ keV, $\mu=0.61$ is the mean molecular weight, and $m_{\rm H}$ is the hydrogen atom mass. We assumed identical IGM conditions for both sources and therefore the same expansion velocity. The dynamical age was then found as $\tau_{dyn}=r_{jet}/v_{exp}$, where $r_{jet}$ is the jet length. For 10913, we find the latter by correcting the observed length of the northern jet in 3 GHz image \citep{3ghz_smol_a} for projection effects using constraints on rotation angles found in \citet{Vulic2025}, and approximating the jet as a straight structure.
This yields a range of intrinsic jet lengths from $\sim330$ kpc (for $\Theta=0^{\circ}$, $\phi=32^{\circ}$) to $\sim390$ kpc ($\Theta=45^{\circ}$, $\phi=53^{\circ}$; see Appendix E in \citealt{Vulic2025} for angle definitions).
For radio galaxy 44, no constraints on the rotation angles are available, and we therefore adopted the projected length measured in the 3 GHz image between the core and the tail tip, $\sim110$ kpc, as the best available estimate for the dynamical age calculation. We additionally consider a plausible range of inclination angles (measured from the line of sight) between $30^{\circ}$ and $80^{\circ}$, resulting in intrinsic tail lengths ranging from $\sim$112 kpc to $\sim$220 kpc, respectively, and discuss how this affects the dynamical age estimate.
We assumed relative uncertainties of 10\% on both the expansion velocity and the jet lengths.
\vspace{-0.3cm}
\subsubsection{Results}
The derived dynamical age of 10913 ranges from $420\pm60$ Myr (for intrinsic jet length $\sim330$ kpc, see above) to $700\pm100$ Myr (for $\sim390$ kpc), while for radio galaxy 44 the projected length of $\sim110$ kpc results in dynamical age of $140\pm20$ Myr. However, the latter is probably higher and could reach $280\pm40$ Myr (at the bottom limit of the assumed inclination angle range).\par
These values are approximately $\sim12$–20 and $\sim7$ (and possibly up to $\sim14$) times larger than the corresponding spectral ages for 10913 and 44, respectively. The dynamical ages are likely overestimated, primarily due to the assumption of a constant expansion speed, whereas the spectral ages derived from BRATS maps (Section~\ref{sec:spectral_age_analysis}) may underestimate the true source ages, most probably due to the mixing of the electron populations. The observed differences are consistent with discrepancy between dynamical and spectral age estimates previously reported in literature \citep{mahatma2020}. We further discuss this is Section~\ref{sec:discussion_specdynage}.

\vspace{-0.3cm}
\section{Discussion}
\label{sec:discussion}
\subsection{Properties and evolution of jet activity}
\label{sec:prop_and_evol_jets}
The properties and evolution of jet activity in the two tailed radio galaxies are constrained by their radio morphology and radio spectral properties. Both galaxies are located within the core region of a massive galaxy group. The group environment has been investigated in previous work \citep{Vulic2025}. An indication of an early group-group merger was found through studying the bent morphology of the WAT radio galaxy 10913. While a detailed discussion of the environment is beyond the scope of the present study, the dynamically active group provides important context for interpreting the observed jet (tail) structure in the two radio galaxies.\par
WAT radio galaxy 10913 displays a well-defined morphology consisting of a core, jets and lobes, with jets that remain well collimated over large distances before terminating in extended lobes (see Section~\ref{sec:tailed_radio_morph}). Such long, highly collimated jets indicate a sustained and stable plasma outflow over the lifetime of the source. The observed asymmetry between the two jets is most likely a consequence of projection effects. The spatially resolved spectral index analysis (Section~\ref{sec:specindmaps}) revealed a systematic steepening of the radio spectrum with increasing distance from the core along the jets, consistent with radiative energy losses experienced by relativistic electrons during transport away from the nucleus. At larger distances, the spectral index flattens within the lobes, as commonly observed in FR II radio galaxies, and generally attributed to the backflow of plasma that has been re-accelerated at the hotspots, namely, the jet termination shocks at the extremities of the source. This observed behavior implies interaction between the jets and the surrounding IGM.
In addition, five distinct regions of enhanced radio brightness, interpreted as hotspots, are identified along the northern jet, with the high-resolution spectral index map (see Appendix~\ref{app:two-frequency-based-specindmaps}) revealing local spectral flattening at these locations. The brightest hotspot is closest to the core and coincides with the point where the jet starts to bend. These features may arise from small shocks associated with changes in local flow conditions or interactions with the surrounding medium.\par
HT radio galaxy 44 has a compact, flat-spectrum core (Section ~\ref{sec:integrated_flux_density_spectra}) and an extended tail (mostly straight but bending slightly toward the north near its outer end), along which the spectral index generally steepens with distance (Section~\ref{sec:specindmaps}), consistent with radiative aging of the transported synchrotron-emitting electrons. An exception to this trend is a region of locally flatter spectral index in the middle of the tail, although the uncertainties in the spectral index are relatively large in this region. However, the feature persists across multiple frequency (dataset) combinations (i.e., see spectral index maps in Section~\ref{sec:specindmaps} and in Appendix~\ref{app:two-frequency-based-specindmaps}). Such local flattening may reflect variations in local flow conditions, possibly due to interaction with the IGM causing reenergization, as previously suggested to be happening in radio galaxies hosted by merging systems \citep{gasperin2017}. However, what we observe here could be affected by the combination of the measurement uncertainties, limited resolution, and projection effects. The tail likely consists of two unresolved jets and is only weakly collimated, as typical of FRI radio galaxies.
\subsection{Spectral versus dynamical age discrepancy}
\label{sec:discussion_specdynage}
For both galaxies, the dynamical ages we derived in Section~\ref{sec:dyn_age_analysis} are considerably larger than the spectral ages we find in Section~\ref{sec:spectral_age_analysis}, namely $\sim12-20$ times for 10913, and at least $\sim7$ (and possibly up to $\sim14$) times larger for radio galaxy 44.
Such discrepancies between spectral and dynamical ages have been previously observed in the literature, and multiple explanations have been proposed \citep{harwood2013, harwood_hardcastle_croston2015, mahatma2020}, although no consensus exists.\par
We derived dynamical ages from the observed source extents under the assumption that the jets (lobes) propagate at a constant speed corresponding to the speed of sound in the surrounding IGM. However, it is likely that lobe advancement decelerates over time due to interaction with the surrounding medium. Analytical (and semi-analytical) models and numerical simulations that aim to reproduce the observed properties of radio sources and their expansion in realistic environments (\citealt{HC2013, TS2015, Hardcastle2018}; see also \citealt{TS2023} for a review) suggest that the assumption of constant lobe advance speed is a significant oversimplification. Employing a more complex dynamical model to properly account for this interaction in our case would require better resolution and sensitivity in X-ray (and multiwavelength) observations of the group and the radio galaxies. Moreover, \citet{mahatma2020} show that even upon employing self-consistent dynamical models that agree with observational constraints (and accounting for sub-equipartition magnetic fields), the age discrepancy of at least a factor of two remains. In conclusion, the simple dynamical model adopted here probably overestimates the true ages of the radio galaxies. This effect could be particularly relevant in dynamically active group environments such as this one, where enhanced interactions between the radio structures and the IGM could further slow the lobe expansion.\par
Conversely, the spectral ages we derived from the BRATS analysis may underestimate the true ages. First, they depend on the assumed lobe magnetic field strength, which is often found under the assumption of equipartition but has been shown in several studies to deviate from this condition \citep{Croston2005, ineson2017}. In this work, we explored both equipartition and non-equipartition scenarios (see Section~\ref{sec:spectral_age_analysis} and Appendix~\ref{app:specage_analysis_details}), and found that the resulting spectral ages are consistent within uncertainties, while the discrepancy with dynamical ages remains. Second, spectral aging analysis is probably affected by the mixing of electron populations with different radiative histories. Re-accelerated plasma in the lobes of the WAT and the tail of the HT galaxy, as previously discussed in Section~\ref{sec:prop_and_evol_jets}, is probably mixing with the older plasma ejected from the radio cores. This can bias the inferred spectral ages towards lower values as previously brought up by \citet{harwood2016}. Additional uncertainties arise from the choice of the injection index and from limitations in resolution, sensitivity, and frequency coverage of the available radio data. Higher-resolution, more sensitive, broad-band radio observations of these galaxies (at both MHz and frequencies higher than 3 GHz) would improve the reliability of spectral age maps and could reduce the discrepancy between spectral and dynamical age estimates. However, such improvements would not eliminate the effects of electron mixing, which remain an inherent limitation of this method.

\subsection{Feedback from the tailed radio galaxies}
Radio galaxies are important drivers of mechanical, or radio-mode, feedback in galaxy groups or clusters, as their radio plasma interacts with the surrounding IGM or ICM \citep{Harrison2017, Wylezalek_morganti2018}. The tailed radio morphologies of the WAT radio galaxy 10913 and HT galaxy 44 studied here clearly indicate ongoing interactions between the jets, lobes, and the IGM. Through these interactions, energy carried by the radio jets is distributed over large regions of the group core, rather than remaining confined close to the host galaxies.\par
The age estimates derived in this work indicate that the radio activity in both galaxies persists for tens to hundreds of Myr, providing sufficient time for such interactions to influence the surrounding IGM. The combination of extended radio structures, spectral index distributions consistent with radiative aging and localized particle re-acceleration, long activity timescales inferred from both spectral and dynamical ages, and clear morphological signatures of jet-IGM interaction indicates that both galaxies are actively interacting with, and probably mechanically depositing energy into their environment.\par
A quantitative estimate of the jet mechanical power would require deep X-ray observations capable of detecting cavities and shocks in the IGM, which are not yet available for this system. While empirical relations between radio luminosity and jet power exist, such estimates are known to carry substantial uncertainties and scatter \citep{birzan2004,cavagnolo2010,godfrey2016}. For this reason, we do not attempt to derive jet powers here. Future deep X-ray and broad-band radio observations will be essential to better constrain the energetics of this interaction and to study its impact on the evolution of the galaxy group.

\vspace{-0.4cm}
\section{Summary}
\label{sec:summary}
In this work, we investigated radio spectral properties and aging of two tailed radio galaxies (a WAT 10913, and HT galaxy 44) located in the core of a massive, dynamically young galaxy group in the COSMOS field at z=0.35. The group was previously detected in X-rays and identified as an early-stage group-group merger. The key findings of this work are summarized below.\par
The tailed radio morphologies reveal clear signatures of interaction with the IGM. The WAT galaxy 10913 exhibits long, well-collimated and bent jets terminating in extended lobes, while the HT galaxy 44 shows a compact core and an extended tail that bends slightly near its end, consistent with ram-pressure effects in the group environment. Spectral index analysis based on integrated radio flux densities at five frequencies between 325 MHz and 3 GHz yielded spectral indices of $\alpha=0.8\pm 0.1$ for 10913 and $\alpha=0.6\pm 0.2$ for 44. Integrated spectra of individual source components, together with spatially resolved spectral index maps show systematic steepening of the radio spectrum with distance from the core in both galaxies, consistent with radiative aging of synchrotron-emitting electrons. Localized spectral flattening, indicative of particle re-acceleration, is observed in the lobes and along the jet of 10913, and in a region of the tail of galaxy 44, although with larger uncertainties.\par
Our dynamical ages are significantly larger than the spectral ages we found from the BRATS analysis. For radio galaxy 10913, we find $\tau_{dyn}=420\pm60$ to $700\pm100$ Myr, compared to a spectral age of $33.80\substack{+7.63 \\ -7.23}$ Myr, while for 44 we obtain $\tau_{dyn}=140\pm20$ Myr (and possibly up to $280\pm40$ Myr) and a spectral age of $20.86\substack{+10.07 \\ -17.17}$ Myr, yielding dynamical-to-spectral age ratios of $\sim12-20$ and $\sim7$ (and possibly up to $\sim14)$, respectively. This discrepancy is likely caused by the use of a simplified dynamical age model that does not account for probable deceleration of the radio lobes due to interaction with the IGM, together with an underestimation of spectral ages arising from mixing of electron populations and limitations in resolution and frequency coverage.
The combination of extended radio structures, spectral signatures of radiative aging with localized re-acceleration, activity timescales of tens to hundreds of Myr, and clear morphological evidence of jet–IGM interaction indicates that both galaxies are actively interacting with, and probably mechanically depositing energy into their environment.
Future deep X-ray and broad-band radio observations will be essential to better constrain the energetics of the feedback and study its role in the evolution of this galaxy group.

\begin{acknowledgements}
P.V. acknowledges support from the project “Implementation of cutting-edge research and its application as part of the Scientific Center of Excellence for Quantum and Complex Systems, and Representations of Lie Algebras”, Grant No. PK.1.1.10.0004, co-financed by the European Union through the European Regional Development Fund - Competitiveness and Cohesion Programme 2021-2027.\\
P.V. gratefully acknowledges the late Branka Paar for her constant support during this work.
\end{acknowledgements}

\bibliographystyle{aa}
\bibliography{bibliography}


\begin{appendix}

\section{Comparison with catalog flux densities}
\label{app:comparison_fluxes}
\begin{figure}[h!]
    \centering
    \includegraphics[width=0.85\hsize]{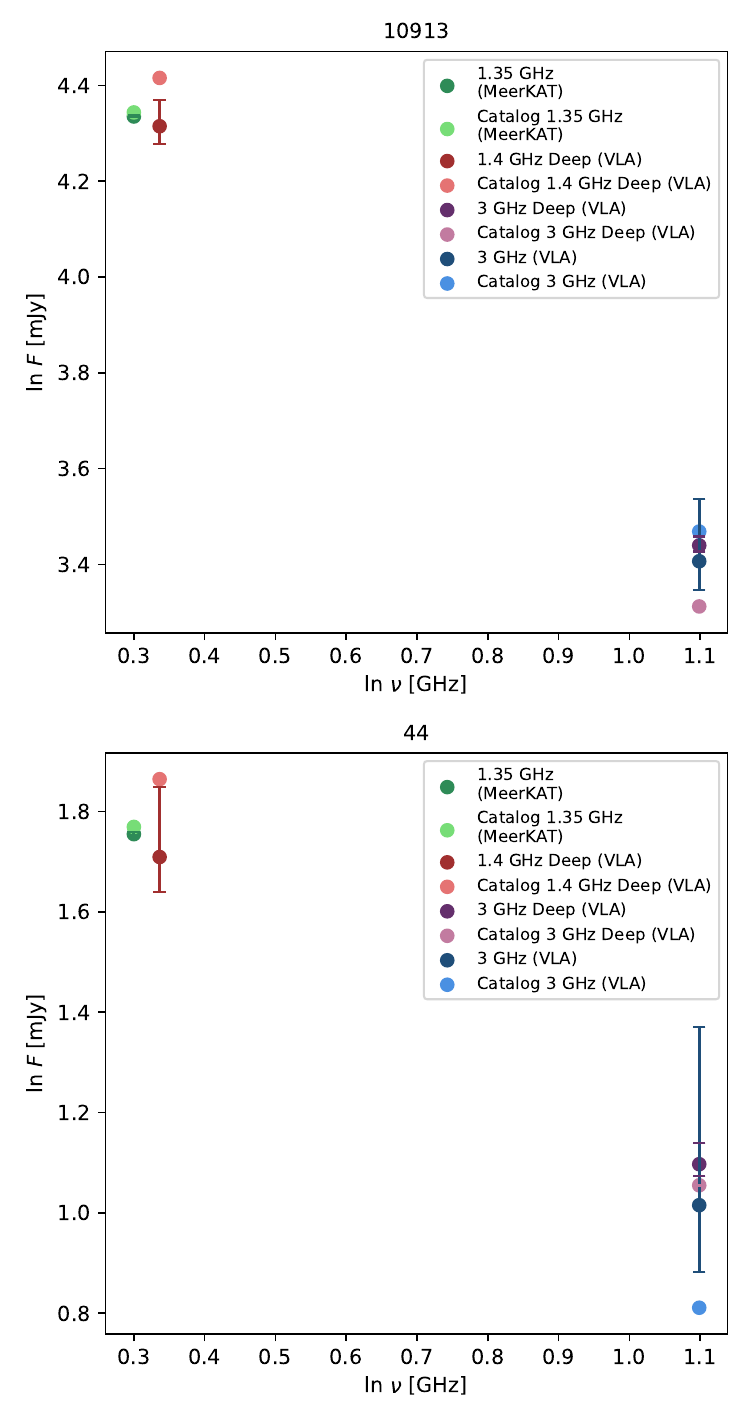}
    \caption{Visual comparison of integrated radio flux densities derived in this work (Sect.~\ref{sec:integrated_flux_density_spectra}) and corresponding catalog values at available frequencies on $\ln-\ln$ scale for radio galaxies 10913 (top) and 44 (bottom). Differences at a given frequency arise from methodological differences between this work and the catalog analyses.}
    \label{fig:comparison_of_fluxes_with_catalog}%
\end{figure}

Fig.~\ref{fig:comparison_of_fluxes_with_catalog} presents a visual comparison between the integrated flux densities of our tailed radio galaxies 10913 (top) and 44 (bottom) derived in this work and the corresponding values from published radio source catalogs, where the latter was available, namely only at 3 GHz and 3 GHz Deep (VLA), 1.4 GHz Deep (VLA) and 1.35 GHz (MeerKAT). The agreement between our measurements and the catalog values is generally good, with relative differences typically below $10\%$, consistent with the uncertainties reported and attributable to differences in the measurement methods. Errors for catalog flux densities were not always available. The largest relative difference ($\sim 18\%$) occurs for radio galaxy 44 at 3 GHz when compared to the value reported by \citet{3ghz_smol_a}, and arises because a substantial fraction of the low-surface-brightness emission in the tail lies between our $3\sigma$ threshold and the $5\sigma$ limit used in the catalog analysis.
\section{Components of tailed radio galaxies}
\label{app:components_of_tailed_rgs}
\subsection{Component identification}
\begin{figure}[h!]
    \centering
    \includegraphics[width=\hsize]{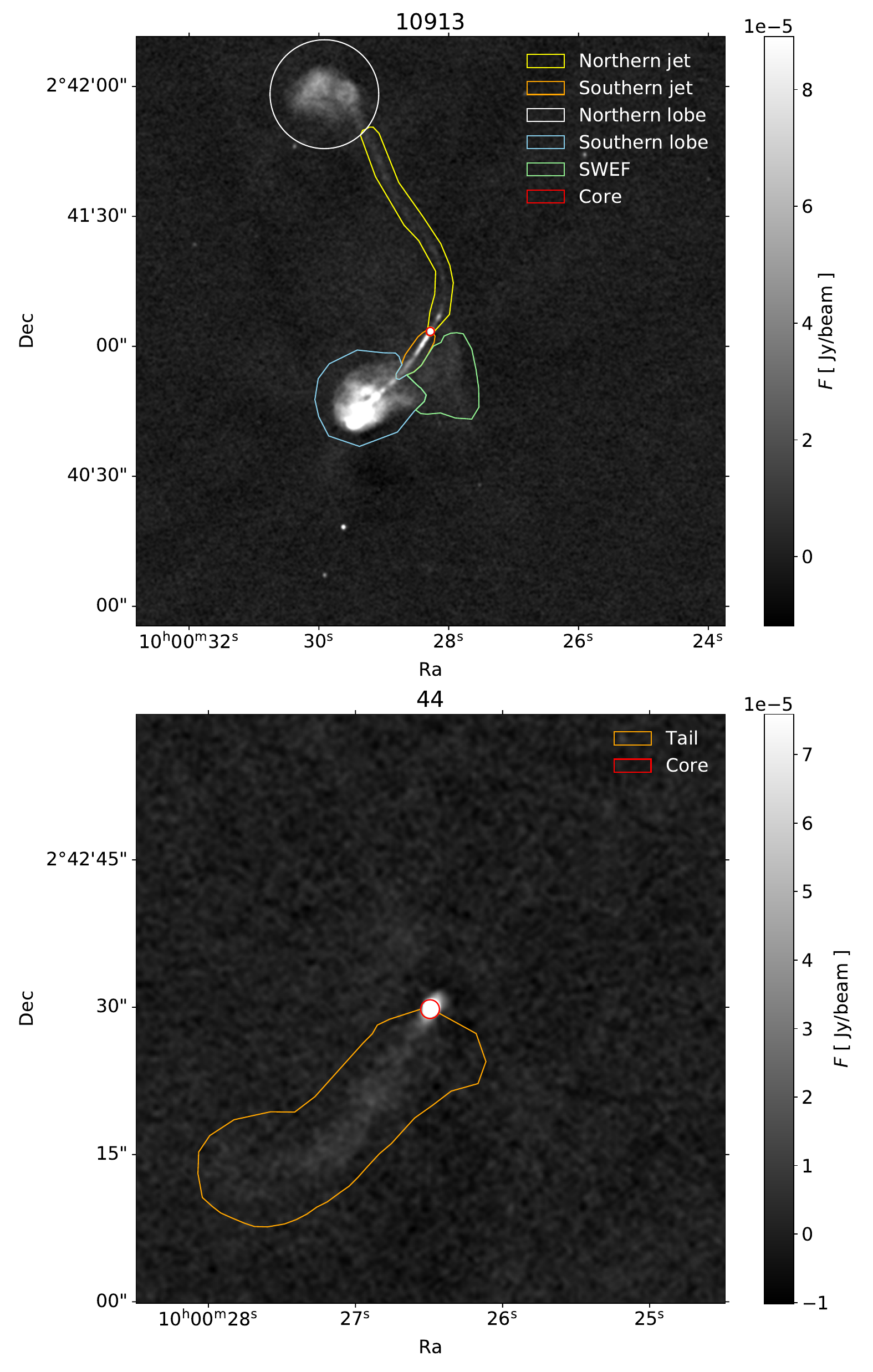}
    \caption{Components of tailed radio galaxies 10913 (top) and 44 (bottom) - regions plotted over VLA radio images at 3 GHz \citep{3ghz_smol_a}. Radio galaxy 10913 is divided into: core, northern and southern jet, northern and southern lobe, and south-west extended feature (SWEF), and radio galaxy 44: core and tail.}
    \label{fig:plotcomp}%
\end{figure}

A careful identification and separation (see Section~\ref{sec:defining_and_separating_galaxy_comp} below) of components of a radio galaxy is important to properly track their spectral evolution. We divide our tailed radio galaxies into separate components by inspecting the best resolution imaging data available to us \citep[3 GHz map cutout, ][]{3ghz_smol_a}, as presented here in Fig.~\ref{fig:plotcomp}. Galaxy 10913 consists of six components: core, northern and southern jet, northern and southern lobe and south-west extended feature (SWEF). In case of radio galaxy 44 the components are its core and tail, where the tail could correspond to unresolved radio jets of a NAT galaxy. For images at other available radio frequencies see Figs.~\ref{fig:fluxandcont10913} and ~\ref{fig:fluxandcont44}, for radio galaxies 10913 and 44, respectively. In Section~\ref{sec:integrated_flux_density_spectra} of this paper we present radio spectral evolution of integrated component flux densities, and calculate the corresponding spectral indices.
\subsection{Limitations in separating components and implications for integrated flux density calculation}
\label{sec:defining_and_separating_galaxy_comp}
Galaxy components are not always clearly separable, caused by resolution limits, projection effects, or as a combination of both. This needs to be taken into account when calculating integrated flux densities of neighboring components, given their mixing can result in underestimating the value for one component, leading to overestimation for the other one.\par

In each cutout (i.e., at each frequency), we first fit a 2D Gaussian to the central (core) region keeping $\sigma_{x}$,$\sigma_{y}$, and the angle of inclination $\theta$ of the Gaussian fixed to the known values of the corresponding Gaussian beam. We do so to be able to deblend (separate) the unresolved core from the rest of the radio structure before the further analysis. The resulting 2D Gaussians cut at $3\sigma$ in both directions, i.e. containing $\approx 99.7\%$ of the Gaussian beam are then considered as the corresponding core regions. 

Then, we define a parameter of separability $s_{comp}^{\nu}$, for a particular galaxy component at a particular frequency $\nu$ as:
\begin{equation}
\label{eq:s}
s_{comp}^{\nu}=\ln(N_{comp}(\nu)/N_{beam}(\nu)).
\end{equation}
Here $N_{comp}(\nu)$ is the number of pixels covering the galaxy component at frequency $\nu$, and $N_{beam}(\nu)$ is the corresponding radio beam size in pixels. This parameter quantifies how well a given component is resolved relative to the beam size at a given frequency. To calculate $s_{comp}^{\nu}$ for different components at available frequencies (from 3 GHz to 325 MHz), we first generate a region file of each component, as found by visually inspecting the best resolution 3 GHz VLA image (Fig. \ref{fig:plotcomp}). Then, for each component we overlay the region on cutouts at other frequencies to find the corresponding numbers of pixels $N_{comp}(\nu)$. We combine this and the matching radio beam size in pixels to obtain $s_{comp}^{\nu}$ using equation \eqref{eq:s}. If $s_{comp}^{\nu}>0$ the component is in principle resolved. Conservatively, if $s_{comp}^{\nu}>1$ we label the component in question as separable (from other galaxy parts) at frequency $\nu$.\par
For each component labeled as separable according to the above definition, we additionally check if all the neighboring components are also labeled as separable at this frequency. If this is met, the corresponding region is used in finding the component's integrated flux density in Section~\ref{sec:integrated_flux_density_spectra}.
Additionally, having found the core regions as explained above, we were able to adapt the regions of the neighboring components (jet(s)/tail) to ensure that there is no overlap or a gap between them.\par 
The above method does not take into account limitations due to projection effects. The latter seems to be affecting the separation of the integrated flux densities of the southern jet and lobe (radio galaxy 10913). The contribution of the jet may be slightly underestimated and the contribution of the lobe overestimated (assuming the region definition we use, Fig.~\ref{fig:plotcomp}).

\section{High-resolution two frequency-based spectral index maps}
\label{app:two-frequency-based-specindmaps}
\begin{figure*}[!h]
    \centering
    \includegraphics[width=0.82\textwidth]{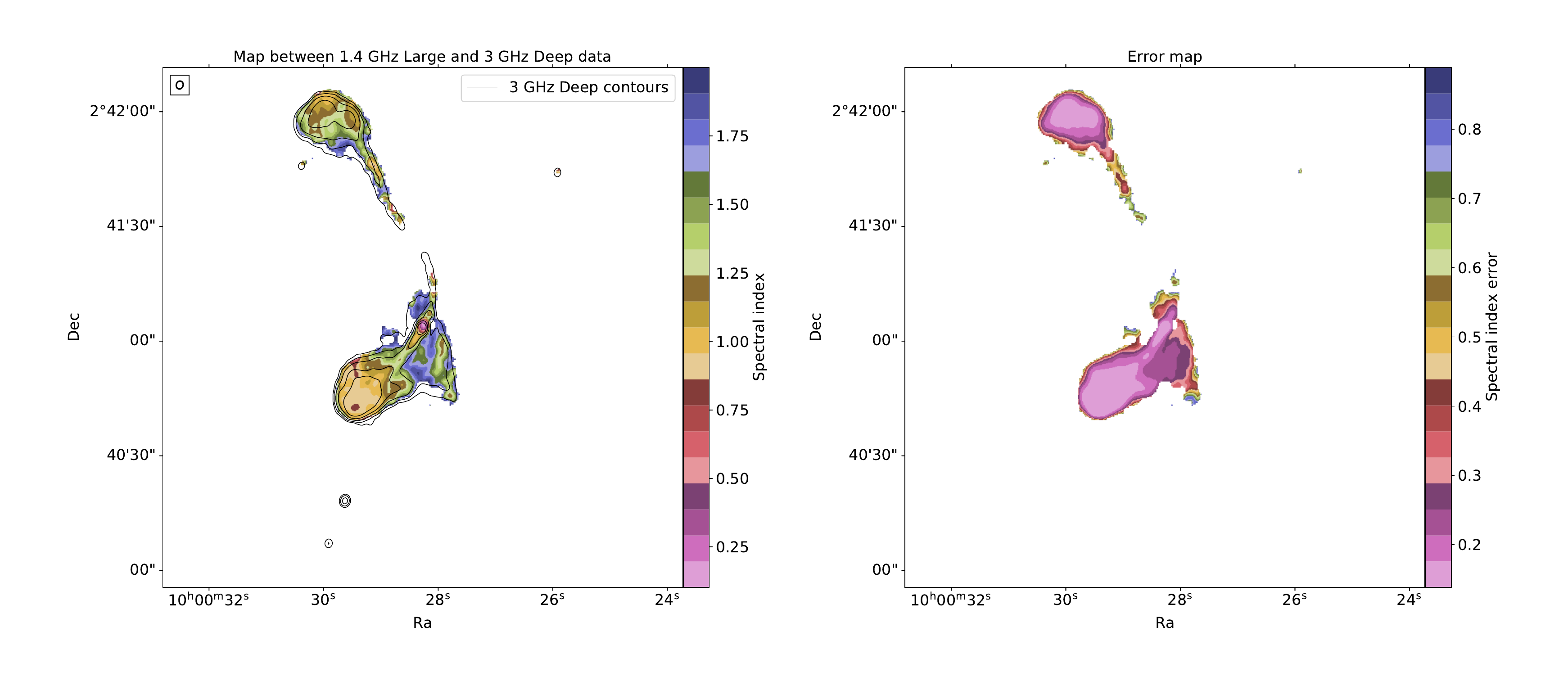}
    \caption{Spectral index map (left) and the corresponding error map (right) for WAT radio galaxy 10913 produced with BRATS using 3 GHz Deep and 1.4 GHz Large datasets. 3 GHz Deep radio contours are overlaid on the spectral index map. Map resolution is set by the 3 GHz Deep data resolution ($\sim 2^{\prime\prime}$) and the corresponding radio beam is shown in the upper left.}
    \label{fig:10913specind}
\end{figure*}

\begin{figure*}[h!]
    \centering
    \includegraphics[width=0.82\textwidth]{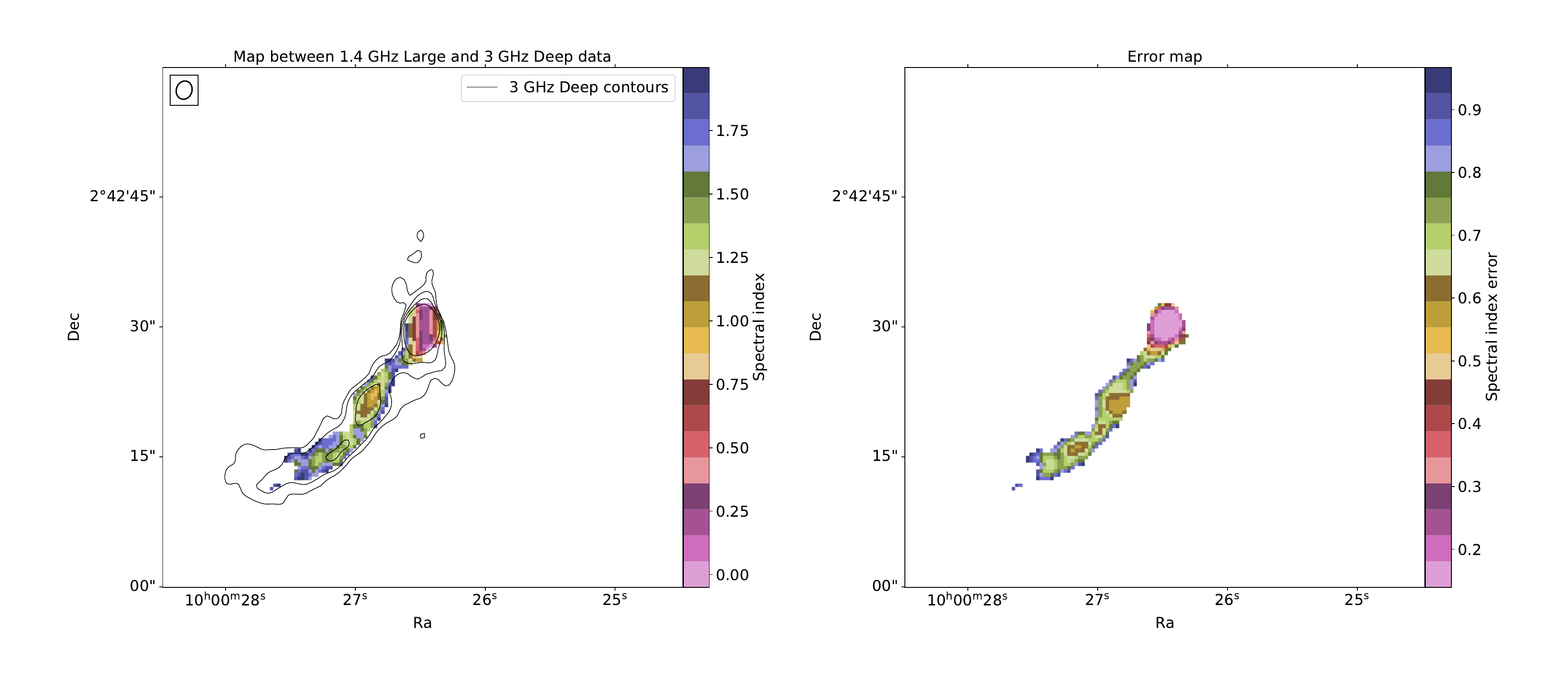}
    \caption{Spectral index map (left) and the corresponding error map (right) for tailed radio galaxy 44 produced with BRATS using 3 GHz Deep and 1.4 GHz Large datasets. See caption of Fig.~\ref{fig:10913specind} for details on radio contours (shown in black) and resolution (radio beam).}
    \label{fig:44specind}
\end{figure*}

We investigate the spatial distribution of spectral index within extended radio sources (galaxies) 10913 and 44 by creating high-resolution pixel-to-pixel spectral index maps and the corresponding error maps, using 3 GHz Deep \citep[VLA;][]{vanderVlugt2021} and 1.4 GHz Large data \citep[VLA;][]{Schinnerer_L}, given their superb depth and resolution, respectively. The maps are created using BRATS software \citep{harwood2013,harwood_hardcastle_croston2015}. For each galaxy, we first smooth the high resolution ($1.4^{\prime\prime}\times 1.5^{\prime\prime}$) 1.4 GHz Large image to the resolution of 3 GHz Deep image ($2^{\prime\prime}$). This is done by using the CASA tool \texttt{imsmooth} (gaussian kernel). Then we regrid the smoothed image to the grid of the lower resolution 3 GHz Deep image using the CASA tool \texttt{imregrid}. Smoothing and regriding resulted in both images having the same elliptical beam $2.14^{\prime\prime}\times 1.81^{\prime\prime}$ and pixel scale of $0.4^{\prime\prime}$. After the image processing in CASA, we checked in DS9 whether the total flux density in a region surrounding each galaxy is the same as in the original images. We find that the values agree within less than 0.1\%. For each source, the radio images were already well aligned astrometrically and any residual offsets were corrected following the procedure described in Section~\ref{sec:specindmaps}, using Gaussian fits to a nearby high S/N point source (for 10913) and to the radio core (for 44). Together with the input images, we load two region files in BRATS, defining the analysis area and the region used for rms noise calculation. The same values of the parameters \texttt{sigma}, \texttt{onsource}, and \texttt{signaltonoise} were adopted as in the three frequency-based analysis (Section~\ref{sec:specindmaps}). We set flux calibration errors to 10\% and 5\% for 3 GHz Deep and 1.4 GHz Large images, respectively. For only two frequencies, the index calculation method reduces to:
\begin{equation}
\alpha=-\frac{\log F_1 - \log F_2}{\log \nu_1 - \log \nu_2},
\end{equation}
where $\nu_1$ and $\nu_2$ are frequencies of two images, and $F_1$ and $F_2$ are the corresponding fluxes of different regions (in our case single pixels). The corresponding errors are calculated by BRATS through error propagation:
\begin{equation}
\Delta \alpha=\log \bigg (\frac{\nu_{1}}{\nu_{2}} \bigg)^{-1}\sqrt{ \bigg(\frac{\Delta F_1}{F_1} \bigg)^{2}+ \bigg(\frac{\Delta F_2}{F_2} \bigg)^{2}}.
\end{equation}
Here, $\Delta F_1$ and $\Delta F_2$ are determined for each pixel by BRATS based on the flux calibration error information that we manually set as described above. The resulting spectral index maps are shown in Figs. \ref{fig:10913specind} and \ref{fig:44specind} for radio galaxies 10913 and 44, respectively, where we additionally overlay the 3 GHz Deep radio contours. In the final maps, we remove a small number of pixels with very high non-physical spectral indices. These were located on galaxy edges and are most probably an artifact of minor misalignment left even after the image preprocessing as described above.


\section{Spectral age analysis with BRATS - Detailed results}
\label{app:specage_analysis_details}
\begin{figure}[!h]
    \centering
    \includegraphics[width=0.45\textwidth]{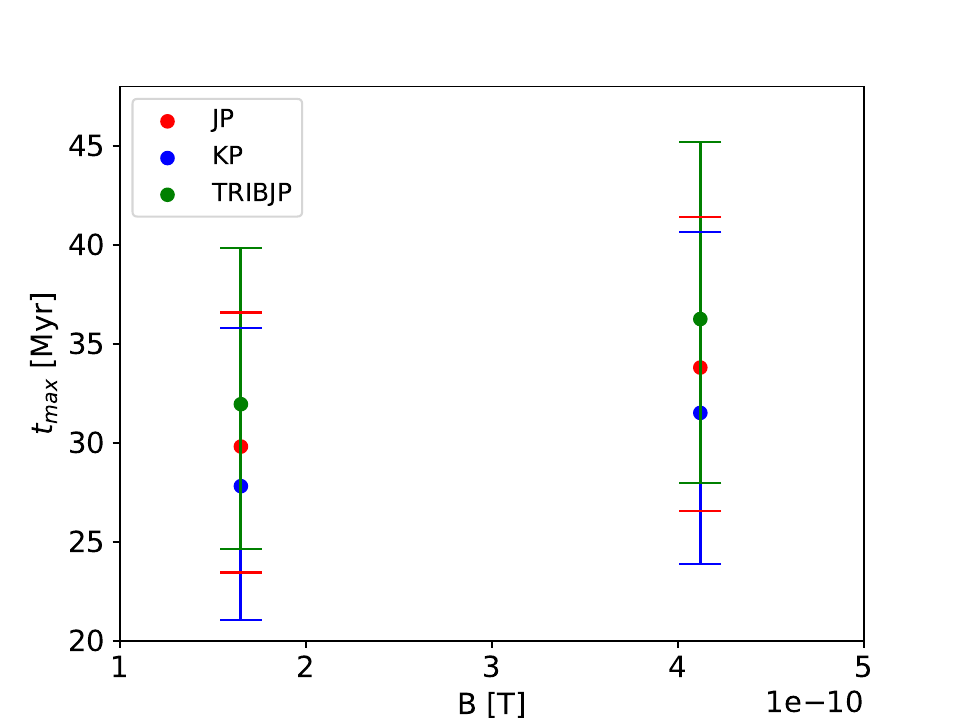}
    \caption{Comparison of the maximum spectral age of WAT radio galaxy 10913 derived from BRATS fits assuming different spectral aging models (JP, KP, TRIBJP) and magnetic field strengths ($B_{eq}$ and $0.4\,B_{eq}$).}
    \label{Fig:modelcomp10913}%
\end{figure}

\begin{figure}[!h]
    \centering
    \includegraphics[width=0.45\textwidth]{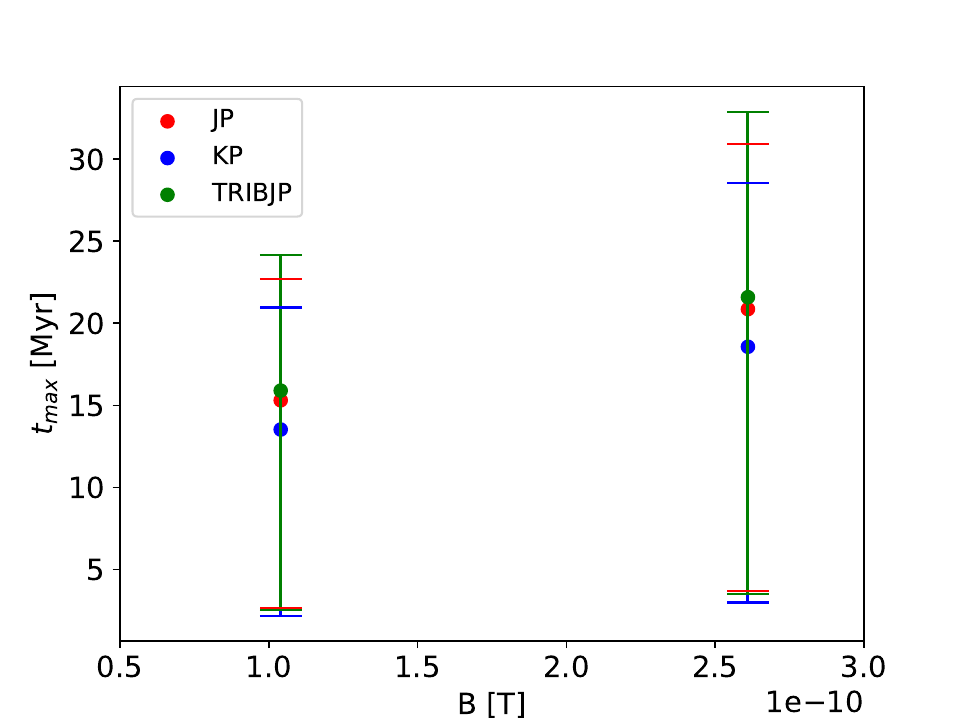}
    \caption{Comparison of the maximum spectral age of tailed radio galaxy 44 derived from BRATS fits assuming different spectral aging models (JP, KP, TRIBJP) and magnetic field strengths ($B_{eq}$ and $0.4\,B_{eq}$).}
    \label{Fig:modelcomp44}%
\end{figure}

We performed a spectral age analysis with BRATS \citep{harwood2013,harwood_hardcastle_croston2015} as described in Section~\ref{sec:spectral_age_analysis}, assuming three different aging models, JP, KP, and TRIBJP and two different lobe magnetic field strengths $B_{eq}$ and $0.4\,B_{eq}$. Here, we compare the resulting ages and the goodness of fit for these six different cases in Tables~\ref{table:BRATSallresults10913} and ~\ref{table:BRATSallresults44}, and Figs.~\ref{Fig:modelcomp10913} and ~\ref{Fig:modelcomp44}.
\newpage
\begin{table*}[]
\caption{Minimum and maximum spectral ages of radio galaxy 10913 derived from BRATS fitting.}
\label{table:BRATSallresults10913}      
\centering          
\begin{tabular}{c | c c c | c c c }    
\hline\hline       
 & \multicolumn{3}{c|}{$B_{eq}=4.12\times10^{-10}\hspace{0.1cm}\mathrm{T}$} & \multicolumn{3}{c}{$0.4\,B_{eq}=1.65\times10^{-10}\hspace{0.1cm}\mathrm{T}$}\\ \hline
 &  JP & KP & TRIBJP & JP & KP & TRIBJP \\ \hline    
  & & & & & & \\                  
 Reduced $\chi^2_{med}$ & 2.408 & 2.455 & 2.458 & 2.410 & 2.455 & 2.458\\
  & & & & & & \\
 $t_{min}$ [Myr] & $0.00\substack{+5.49 \\ -0.00}$ & $0.00\substack{+4.78 \\ -0.00}$ & $0.00\substack{+5.59 \\ -0.00}$ & $0.00\substack{+4.87 \\ -0.00}$ & $0.00\substack{+4.28 \\ -0.00}$ & $0.00\substack{+7.74 \\ -0.00}$ \\
 & & & & & & \\
 $t_{max}$ [Myr] & $33.80\substack{+7.62 \\ -7.23}$ & $31.51\substack{+9.12 \\ -7.61}$ & $36.25\substack{+8.93 \\ -8.30}$ & $29.81\substack{+6.77 \\ -6.36}$ & $27.81\substack{+8.01 \\ -6.77}$ & $31.95\substack{+7.89 \\ -7.32}$\\
 & & & & & & \\
\hline \hline
\end{tabular}
\tablefoot{
Ages are obtained by fitting JP, KP, and TRIBJP spectral ageing models to broad-band radio data at 3 GHz, 1.4 GHz Deep, and 610 MHz. Results are shown for two assumed magnetic field strengths in the radio lobes: the equipartition value $B_{eq}$ and a reduced value equal to $0.4\,B_{eq}$.}
\end{table*}
\begin{table*}[]
\caption{Minimum and maximum spectral ages of radio galaxy 44 derived from BRATS fitting.}             
\label{table:BRATSallresults44}      
\centering          
\begin{tabular}{c | c c c | c c c }    
\hline\hline       
 & \multicolumn{3}{c|}{$B_{eq}=2.61\times10^{-10}\hspace{0.1cm}\mathrm{T}$} & \multicolumn{3}{c}{$0.4\,B_{eq}=1.04\times10^{-10}\hspace{0.1cm}\mathrm{T}$}\\ \hline
 &  JP & KP & TRIBJP & JP & KP & TRIBJP \\ \hline    
  & & & & & & \\                  
 Reduced $\chi^2_{med}$ & 1.602 & 1.607 & 1.608 & 1.602 & 1.606 & 1.608\\
  & & & & & & \\
 $t_{min}$ [Myr] & $3.62\substack{+13.96 \\ -3.62}$ & $3.18\substack{+12.40 \\ -3.18}$ & $3.92\substack{+14.18 \\ -3.92}$ & $2.59\substack{+10.30 \\ -2.59}$ & $2.15\substack{+9.32 \\ -2.15}$ & $2.74\substack{+10.59 \\ -2.74}$ \\
 & & & & & & \\
 $t_{max}$ [Myr] & $20.86\substack{+10.07 \\ -17.17}$ & $18.57\substack{+9.95 \\ -15.57}$ & $21.59\substack{+11.33 \\ -18.10}$ & $15.31\substack{+7.40 \\ -12.67}$ & $13.53\substack{+7.43 \\ -11.33}$ & $15.90\substack{+8.27 \\ -13.38}$\\
 & & & & & & \\
\hline \hline
\end{tabular}
\tablefoot{
Ages are obtained by fitting JP, KP, and TRIBJP spectral ageing models to broad-band radio data at 3 GHz, 1.4 GHz Deep, and 610 MHz. Results are shown for two assumed magnetic field strengths in the tail: the equipartition value $B_{eq}$ and a reduced value equal to $0.4\,B_{eq}$.}
\end{table*}


\end{appendix}
\end{document}